\documentclass[aps,prl,twocolumn,superscriptaddress,amssymb,amsfonts,longbibliography,10pt]{revtex4-2}

\usepackage[latin9]{inputenc}
\usepackage{color}
\usepackage{verbatim}
\usepackage{amsmath}
\usepackage{amssymb}
\usepackage{dsfont}
\usepackage{graphicx}
\usepackage{esint}
\usepackage{esvect}
\usepackage{braket}
\usepackage{amsmath,bm}
\usepackage{makecell}
\usepackage{graphicx}
\usepackage{pifont}
\newcommand{\xmark}{\ding{55}}
\usepackage{subfigure}
\usepackage{url}
\usepackage{lipsum}
\usepackage[dvipsnames]{xcolor}
\usepackage[
    colorlinks,
    linkcolor={blue!80!black},
    citecolor={blue!80!black},
    urlcolor={blue!80!black}
]{hyperref}

\usepackage{gensymb}
\usepackage{xfrac}

\makeatletter
\@ifundefined{textcolor}{}
{%
 \definecolor{BLACK}{gray}{0}
 \definecolor{WHITE}{gray}{1}
 \definecolor{RED}{rgb}{1,0,0}
 \definecolor{GREEN}{rgb}{0,1,0}
 \definecolor{BLUE}{rgb}{0,0,1}
 \definecolor{CYAN}{cmyk}{1,0,0,0}
 \definecolor{MAGENTA}{cmyk}{0,1,0,0}
 \definecolor{YELLOW}{cmyk}{0,0,1,0}
}

\newcommand{\hodge}{\,\boldsymbol{\star}\,}
\newcommand{\bk}{\mathbf{k}}

\newcommand{\bn}{\mathbf{n}}
\newcommand{\bb}{\mathbf{b}}
\newcommand{\bJ}{\mathbf{J}}
\newcommand{\bR}{\mathbf{R}}

\newcommand{\br}{\mathbf{r}}

\newcommand{\be}{\mathbf{e}}
\newcommand{\bh}{\mathbf{h}}
\newcommand{\bS}{\mathbf{S}}
\newcommand{\bd}{\mathbf{d}}

\newcommand{\bM}{\mathbf{m}}

\newcommand{\bo}{\mathbf{o}}

\newcommand{\cP}{\mathcal{P}}
\newcommand{\cT}{\mathcal{T}}
\newcommand{\cK}{\mathcal{K}}

\newcommand{\bsigma}{\boldsymbol{\sigma}}

\newcommand{\blambda}{\boldsymbol{\lambda}}

\usepackage{soul}
\usepackage{ulem}

\draft
\makeatother
\begin{document}

\title{Microscopic origin of $p$-wave magnetism}
\author{Johannes Mitscherling}
\email[]{mitscherling@pks.mpg.de}
\affiliation{Max Planck Institute for the Physics of Complex Systems, N\"othnitzer Str. 38, 01187 Dresden, Germany}
\author{Jan Priessnitz}
\affiliation{Max Planck Institute for the Physics of Complex Systems, N\"othnitzer Str. 38, 01187 Dresden, Germany}
\author{Clara K. Geschner}
\affiliation{Max Planck Institute for the Physics of Complex Systems, N\"othnitzer Str. 38, 01187 Dresden, Germany}
\author{Libor \v Smejkal}
\email[]{lsmejkal@pks.mpg.de}
\affiliation{Max Planck Institute for the Physics of Complex Systems, N\"othnitzer Str. 38, 01187 Dresden, Germany}
\affiliation{Max Planck Institute for Chemical Physics of Solids, N\"othnitzer Str. 40, 01187 Dresden, Germany}
\affiliation{Institute of Physics, Czech Academy of Sciences, Cukrovarnick\'a 10, 162 00, Praha 6, Czech Republic}

\date{\today}

\begin{abstract}
$P$-, $f$-, or $h$-wave antialtermagnets yield large non-relativistic spin splitting with out-of-plane spin polarization in momentum space perpendicular to the coplanar non-collinear local magnetic moments. We provide a microscopic explanation of this unconventional spin polarization by linking it to a previously overlooked site-compensated spin density that orders antiparallel when projected onto opposite momenta. We verify this result both by model derivation of the out-of-plane momentum-space spin polarization being proportional to the direct-space cross product of the coplanar non-collinear spin order, as well as by ab initio calculations in the material candidate CeNiAsO. By providing a general classification and analytic expression for the spin polarization of all spinful two-site tight-binding Hamiltonians, we reveal the momentum-resolved spin polarization as a probe of the Bloch-state geometry arising from spin-site coupling. Furthermore, our approach allows for geometric distinction between ferro-, alter-, and antialtermagnets. Our results provide a quantitative guidance for quantized out-of-plane momentum-space spin polarization and large spin splitting, and construction principles for antialtermagnets.
\end{abstract}

\maketitle

\begin{figure}
    \centering
    \includegraphics[width=0.99\linewidth]{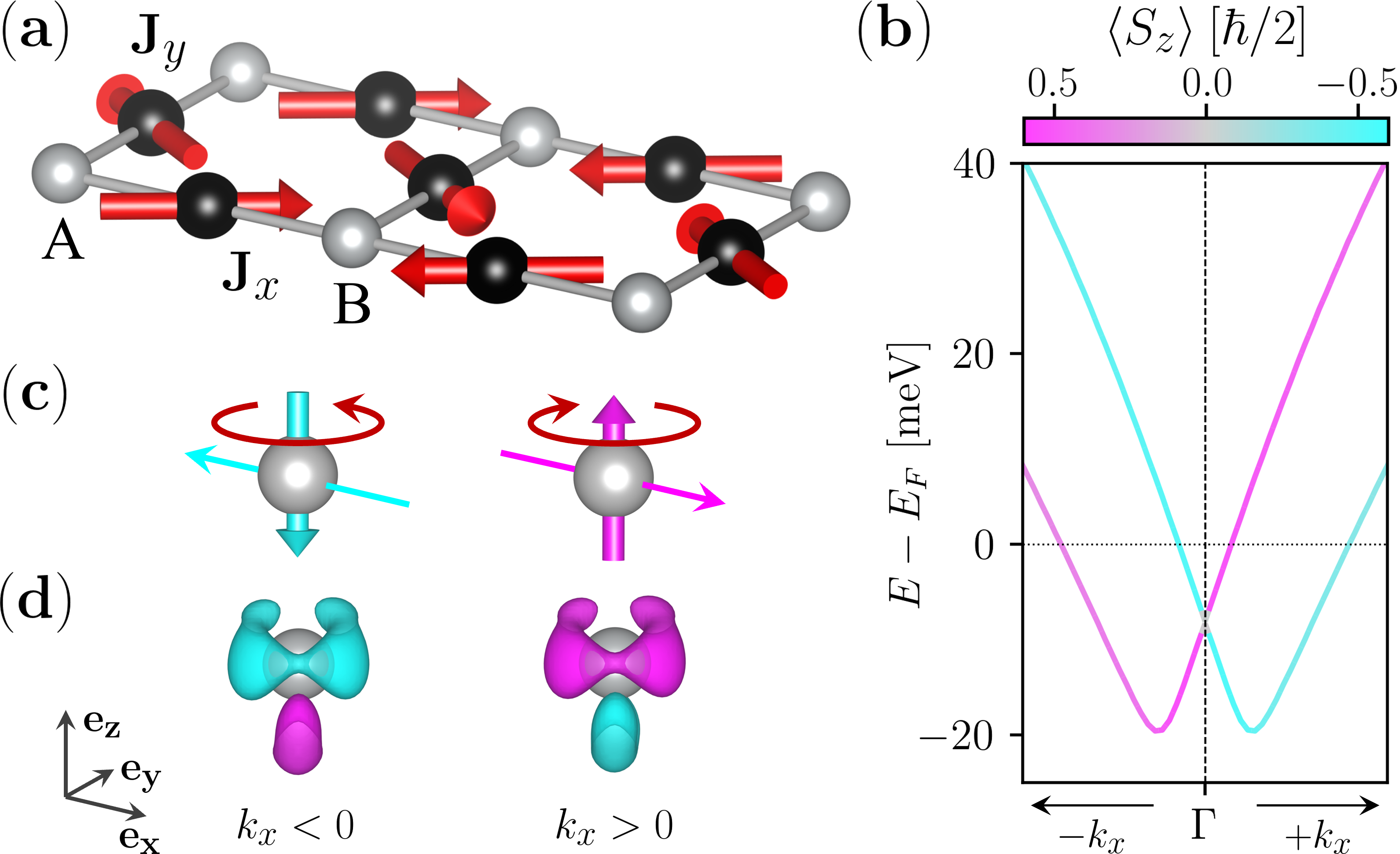}
    \caption{Schematics of a coplanar magnetic order (a) that yields non-relativistic band splitting with out-of-plane $p$-wave spin polarization (b). The band-projected spin polarization perpendicular to the in-plane magnetic moments is proportional to their cross product felt by forward and backward propagating electrons (c), arising from Bloch-state geometry. The hidden antiparallel real-space spin density becomes visible after directional selection of the electrons (d).}
    \label{fig:Model}
\end{figure}

{\it Introduction.---} Recently, odd-parity-wave magnets~\cite{Hellenes2024} have been proposed by employing spin group theory, previously used to identify and classify altermagnets~\cite{Smejkal2022}. In contrast to even-parity $d$-, $g$-, and $i$-wave altermagnets~\cite{Smejkal2022b}, the odd-parity $p$-, $f$-, and $h$-wave magnets preserve time-reversal symmetry in momentum space and exhibit odd-in-momentum nonrelativistic exchange spin splitting~\cite{Hellenes2024, Brekke2024, Yu2025}. The splitting arises from compensated coplanar non-collinear magnetic moments constrained by a combined time-reversal and lattice-translation symmetry~\cite{Hellenes2024}; see Fig.~\ref{fig:Model}(a,b) for the schematics and spin-polarized band splitting of a minimal $p$-wave model. Various properties arising from the unconventional spin-polarized band structure have been predicted. These include transport anisotropy~\cite{Hellenes2024, Ezawa2025a}, nonrelativistic current-induced spin polarization (Edelstein effect)~\cite{Chakraborty2025, Pari2025}, Type-II unconventional magnetic multiferroics~\cite{Priessnitz2026}, chiral magnons whose spin polarization is odd in momentum~\cite{Neumann2026}, superconducting phenomena~\cite{Sukhachov2025, Fukaya2025}, and Ising superconductivity~\cite{Khodas2026}.

A broad set of material candidates has been identified~\cite{Hellenes2024}, including $h$-wave symmetry~\cite{Yu2025}. Recently, certain $p$-wave characteristics have been confirmed experimentally in incommensurate $\text{NiI}_2$~\cite{Song2025}, rhodium-doped $\text{Gd}_3(\text{Ru}_{1-\delta}\text{Rh}_\delta)_4\text{Al}_{12}$~\cite{Yamada2025}, and $\text{CeNiAsO}$~\cite{Zhou2025}, showing switchable photocurrents~\cite{Song2025}, anomalous Hall response due to a secondary weak magnetization~\cite{Yamada2025}, and transport anisotropy~\cite{Yamada2025, Zhou2025}. While the proportionality of the odd-parity spin splitting to the cross product of the magnetic moments has recently been pointed out~\cite{Yu2025}, the evaluation of the spin polarization and a Bloch-state geometric classification has been lacking. Furthermore, while altermagnets have both nontrivial alternating momentum- and ferroic real-space order~\cite{Smejkal2022b}, the real-space order complementing the out-of-plane collinear order in momentum space of $p$-wave magnets has yet to be identified.

\begin{figure*}
    \centering
    \includegraphics[width=1\linewidth]{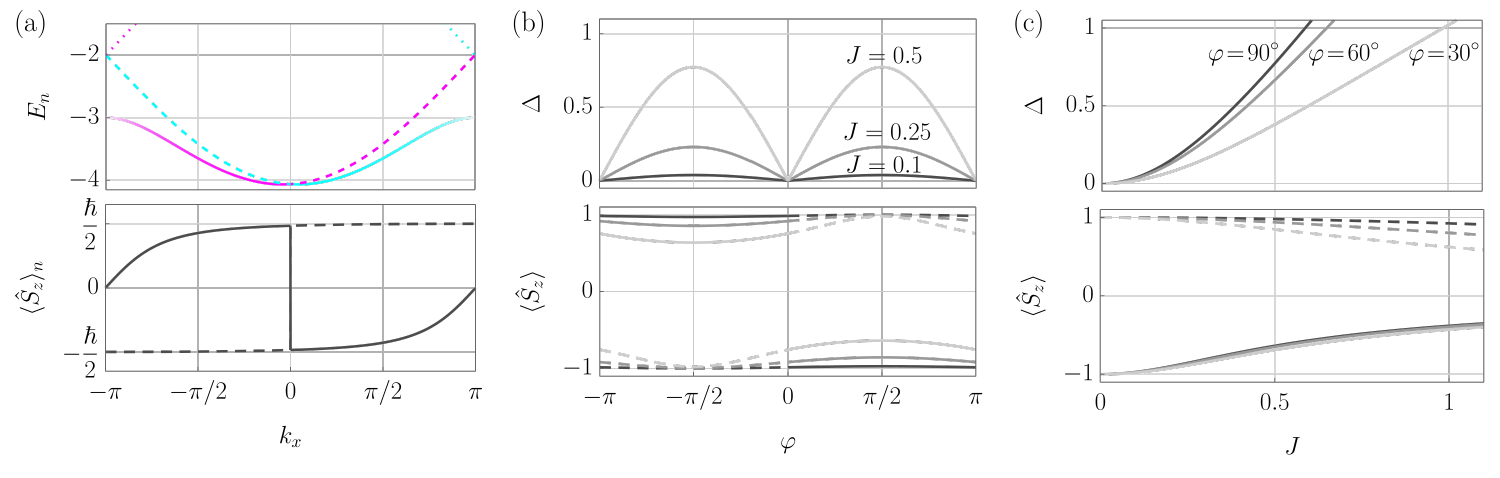}
    \caption{(a) Momentum-dependence of the bands $E_1$ and $E_2$. The spin polarization $\langle \hat S_z\rangle_n$ of the lower (solid) and upper (dashed) band is $\cT$-even and nearly quantized to $\pm\hbar/2$ near $\Gamma=(0,0)$. (b) The spin splitting $\Delta$ yields a sinusoidal angle dependence, maximal at orthogonal in-plane magnetic moments. In contrast, the spin polarization shows a weak angular dependence that increases with the amplitude $J$. (c) The splitting increases quadratically with the amplitude of the in-plane magnetic moments, while the quantization of the spin polarization is gradually lifted, as shown for different relative angles.}
    \label{fig:ParameterDependence}
\end{figure*}

In this work, we close this gap by providing a microscopic real-space perspective on $p$-wave magnets, clarifying the Bloch-state geometric origin of the out-of-plane spin polarization and revealing an antiparallel, compensated intra-site spin order. In particular, we show that the spin-site coupling arising from the non-collinear magnetic moments induces the spin polarization proportional to their cross-product,
\begin{align}
    \langle \hat \bS\rangle_n(\bk) \propto -\,\text{sign}\big[k_x\big]\frac{\bJ_x\times\bJ_y}{|\bJ_x\times\bJ_y|} = \text{sign}\big[k_x\,\sin\varphi\big]\,\be_z \, ,
    \label{eqn:OutOfPlanePolarization}
\end{align}
sensitive to the relative angle $\varphi$ between the in-plane magnetic moments $\bJ_x$ and $\bJ_y$ and the forward and backward propagation via momentum $k_x$; see Fig.~\ref{fig:Model}(c). Using the expansion of the analytic band projector in terms of {\it su}(4) generators~\cite{Graf2021}, we obtain a closed analytic form of the spin polarization of all spinful two-site tight-binding Hamiltonian, enabling a Bloch-state geometric classification of ferro-, alter-, and $p$-wave magnets in terms of the {\it su}(4) star product that captures the anticommutativity of the algebra. We find qualitative agreement between our phenomenological theory and ab initio calculations for the $p$-wave material candidate $\text{CeNiAsO}$. In particular, by distinguishing forward and backward propagating Bloch electrons by momentum projection to the $k_x>0$ and $k_x<0$ half of the Brillouin zone, we reveal an antiparallel real-space spin polarization of the band-projected Kohn-Sham orbitals; see Fig.~\ref{fig:Model}(d).

{\it Spin polarization of $p$-wave magnets.---} We illustrate the relation between the local in-plane magnetic moments and the out-of-plane spin polarization for the minimal two-dimensional four-band model \cite{Hellenes2024}
\begin{align}
    \hat H(\bk) &=  2\,t\big[\cos k_y+\cos (k_x/2)\,\hat\tau_1\big] \nonumber\\ &+2\,\big[\cos k_y\,\bJ_y\,\hat \tau_3-\sin(k_x/2)\,\bJ_x\,\hat\tau_2\big]\cdot\hat \bsigma \, ,
    \label{eqn:PwaveBlochHamiltonian}
\end{align}
where we include spin-independent nearest-neighbor hopping $t$ and spin- and directional-dependent hopping $\bJ_a$ between sites A and B; see Fig.~\ref{fig:Model}(a). We denote the Bloch Hamiltonian in terms of the Kronecker product $\hat \tau_a\hat \sigma_b$, where $\hat \tau_a$ and $\hat \bsigma=(\hat\sigma_1,\hat\sigma_2,\hat\sigma_3)$ are the Pauli matrices for the site and spin space, respectively. We omit the unit matrices $\hat 1_2$ for shorter notation. We assume in-plane magnetic moments of equal strength $|\bJ_a|=J>0$ and characterized by angles $\varphi_a$, i.e., $\bJ_a/|\bJ_a|= \cos\varphi_a\,\be_x+\sin\varphi_a\,\be_y$. The $\cP\cT$-symmetry is broken by $\bJ_x$, lifting spin degeneracy except at isolated momenta. In the following, we focus on the two lowest bands $E_1(\bk)$ and $E_2(\bk)$. The corresponding band- and momentum-resolved spin polarization reads $\langle \hat \bS\rangle_n(\bk) = \text{tr}\big[\hat \bsigma\,\hat P_n(\bk)\big]$ with corresponding orthogonal rank-1 band projectors $\hat H(\bk)\hat P_n(\bk) = E_n(\bk)\hat P_n(\bk)$ and trace over the $4\times4$ matrix structure. In Fig.~\ref{fig:ParameterDependence}(a), we show the band dispersions and corresponding spin polarization as a function of $k_x$ for $J/|t|=0.25$ and orthogonal magnetic moments with relative angle $\varphi\equiv \varphi_x-\varphi_y=\pi/2$. We fix $t=-1$ and $k_y=0$.

As presented in detail in the Supplemental Material (SM)~\cite{SupplMat} and outlined for the general Hamiltonian in the next section, we obtain analytic expressions for the band dispersions and spin polarizations, revealing the full parametric dependence on the angle $\varphi$ and strength $J/|t|$ of the in-plane magnetic moments. The spin splitting $\Delta(\bk) = E_1(\bk)-E_2(\bk)$ yields a strong angle dependence, vanishing for collinear order and becoming maximal for orthogonal magnetic moments; see Fig.~\ref{fig:ParameterDependence}(b) for $\bk=(\pi/2,0)$. Expanding the full expression (see End Matter) to leading order in momentum near $\Gamma = (0,0)$, we obtain
\begin{align}
    \Delta(\bk) \approx \frac{2J^2}{\sqrt{t^2+J^2}}|k_x\,\sin\varphi| \, ,
    \label{eqn:BandGapMainText}
\end{align}
tied to the cross product between the in-plane magnetic moments, $\bJ_x\times\bJ_y\propto \sin\varphi$~\cite{Yu2025}. The gap yields a crossover from $\Delta\propto J^2$ for weak to $\Delta\propto |J|$ for large magnetic order; see Fig.~\ref{fig:ParameterDependence}(c) for $\bk=(\pi/2,0)$.

The spin polarization of the lowest band near $\Gamma$ reads
\begin{align}
    &\langle \hat \bS\rangle_{1}(\bk) \approx  \frac{t\,\text{sign}\big[k_x\,\sin\varphi\big]}{\sqrt{t^2+J^2}}\bigg[1-\frac{1}{2}\frac{J^2}{t^2+J^2}|k_x\,\sin\varphi|\bigg]\be_z\, , 
    \label{eqn:MomentumDependenceSpinPolarization}
\end{align}
which reveals fully polarized, uniform spin polarization in the weak-field limit $J/|t|\ll 1$. In this limit, the spin sectors of the Hamiltonian decouple, allowing for an effective two-band description~\cite{Khodas2026}. Up to chirality, the spin quantization is insensitive to the angle $\varphi$ between the in-plane magnetic moments to leading order; see Fig.~\ref{fig:ParameterDependence}(b). The quantization is lifted for increasing $J/|t|$, being least sensitive for the orthogonal spin pattern and differing between the two lowest bands; see Fig.~\ref{fig:ParameterDependence}(c).

\begin{table}
    \centering
    \begin{tabular}{cccccc}
        \hline\hline\\[-3mm]
        & \hspace{1mm}$\cP_x$\hspace{1mm} & \hspace{1mm}$\cT$\hspace{1mm} &  \hspace{1mm}$\cP_x\cT$-even\hspace{1mm} & \hspace{1mm}Coefficient\hspace{1mm} &$p$-wave magnet~\cite{Hellenes2024} \\ \hline\\[-3mm]
        $\hat 1$     &$+$& $+$ & $\hat 1$& $d_0(\bk)$   &   $2t\cos k_y$   \\\hline\\[-3mm]
        $\hat \tau_1$       &$+$&$+$& $\hat\Gamma_{1}$ & $d_1(\bk)$    &  $2t\cos(k_x/2)$     \\
        $\hat\tau_2$       &$-$&$-$                                       & $\hat\Gamma_{2}$                                     &  $d_2(\bk)$                                     &-                                                           \\
        $\hat\tau_3$       &$-$&$+$                                    &      & $d_3(\bk)$& -                                                                     \\\hline\\[-3mm]
        $\hat \bsigma$ &$+$&$-$&  & $\bd_0(\bk)$ &  -      \\ 

        $\hat\tau_1\hat \bsigma$   &$+$&$-$   &   & $\bd_1(\bk)$                                   &  -                                                          \\ 

        $\hat\tau_2\hat \bsigma$   &$-$&$+$         &  & $\bd_2(\bk)$ &                                 $-2\sin(k_x/2)\,\bJ_x$                            \\
        $\hat\tau_3\hat \bsigma$  &$-$&$-$   & $(\hat\Gamma_{3},\hat\Gamma_{4},\hat\Gamma_{5})$   &  $\bd_3(\bk)$    &  $2\cos(k_y)\,\bJ_y$                             \\[0mm] \hline\hline
    \end{tabular}
    \caption{The decomposition of a general spinful two-site Hamiltonian into the spin-independent $d_a(\bk)$ and spin-dependent functions $\bd_a(\bk)$ with their respective symmetry behaviors, coefficients, and functional form of a minimal $p$-wave model.}
    \label{tab:SummaryMinimalModelPwaveMainText}
\end{table}

{\it General spinful two-site Hamiltonian.---} The specific results for the minimal model in Eq.~\eqref{eqn:PwaveBlochHamiltonian} suggest a strong interplay between {\it real-space} magnetic moments and the geometry of the {\it complex projective state space} of the band electrons, i.e., the Bloch-state geometry, that becomes accessible via the band-projected spin polarization. To make this connection precise, we consider the general spinful two-site Hamiltonian 
\begin{align}
    \hat H(\bk) &= d_0(\bk)+\!\sum_{a=1}^3 d_a(\bk)\hat \tau_a +\bd_0(\bk)\!\cdot\!\hat\bsigma+\!\sum_{a=1}^3 \bd_a(\bk)\cdot \hat\tau_a\hat\bsigma  ,
    \label{eqn:BlochHamiltonian}
\end{align}
with 16 momentum-dependent coefficients within the basis $\hat c^\dagger_\bk = \big(\hat c^\dagger_{A\uparrow}(\bk),\hat c^\dagger_{A\downarrow}(\bk),\hat c^\dagger_{B\uparrow}(\bk),\hat c^\dagger_{B\downarrow}(\bk)\big)$ for the creation operators. We combined the spin-related coefficients into vectors $\bd_a(\bk)$, assuming that the spin frame $\hat \bsigma = (\hat \sigma_1,\hat \sigma_2,\hat \sigma_3)$ is aligned with the right-handed orthogonal spatial coordinate system given by $\be_x$, $\be_y$, and $\be_z$. The behavior under parity $\cP_x=\hat \tau_1$ and time-reversal $\cT$ symmetry, as well as the relation to the minimal $p$-wave model in Eq.~\eqref{eqn:PwaveBlochHamiltonian}, is summarized in Tab.~\ref{tab:SummaryMinimalModelPwaveMainText}.

The presence of $\cP\cT$ symmetry enforces doubly degenerate bands. For instance, the five $\cP_x\cT$-symmetric Dirac matrices $\hat\Gamma_a$, see Tab.~\ref{tab:SummaryMinimalModelPwaveMainText}, satisfy the anticommutation relation $\{\hat\Gamma_a,\hat\Gamma_b\}=2\delta_{ab}$ that implies the doubly degenerate bands $d_0(\bk)\pm\sqrt{d_1(\bk)^2+d_2(\bk)^2+|\bd_3(\bk)|^2}$. The band degeneracy is lifted by breaking the symmetry, for instance, via $\bd_2(\bk)$, leading to 
\begin{align}
    &E_\pm(\bk)\!=\!\sqrt{d_1(\bk)^2\!+\!|\bd_2(\bk)|^2\!+\!|\bd_3(\bk)|^2\!\pm\!2\,|\bd_2(\bk)\!\times\!\bd_3(\bk)|}\, 
    \label{eqn:GeneralBandSplitting}
\end{align}
for the two lower bands $E_{1,2}(\bk) = d_0(\bk)-E_\pm(\bk)$. Eq.~\eqref{eqn:GeneralBandSplitting} reveals that the band splitting in Eq.~\eqref{eqn:BandGapMainText} is generically tied to the cross product of the spin-dependent (vector-valued) site-exchange and imbalance terms $\bd_2$ and $\bd_3$. Note that the origin of the spin splitting fundamentally differs from those of minimal ferro- and altermagnetic four-band models, where the splitting is proportional to $|d_1(\bk)\bd_0(\bk)|$ and $|d_3(\bk)\bd_3(\bk)|$, respectively; see SM~\cite{SupplMat}.

{\it Connection of spin-space and {\it su}(4) algebra.---} The analytic form of the band dispersion, such as in Eq.~\eqref{eqn:GeneralBandSplitting}, enables analytic expressions of the band projectors by employing Lie algebra techniques~\cite{Graf2021}. For this, we choose the orthogonal basis $\hat \blambda = (\hat\tau_i,\hat\bsigma,\hat\tau_i\hat\bsigma)$ with compact notation $\hat \tau_i = (\hat \tau_1,\hat\tau_2,\hat\tau_3)$ for traceless Hermitian $4\times 4$ matrices and expand the general Hamiltonian in Eq.~\eqref{eqn:BlochHamiltonian} as $\hat H(\bk) = d_0(\bk)+\bh(\bk)\cdot\hat \blambda$ with Hamiltonian vector $\bh(\bk) = \big(d_i(\bk),\bd_0(\bk),\bd_i(\bk)\big)$. Similarly, we expand the spin operator as $\hat S_a = \bar\be_a\cdot\hat \blambda$, where $\bar\be_a = (\mathbf{0}_3,\be_a,\mathbf{0}_9)$ for $a=x,y,z$ is the embedding of the coordinate vector into the 15-dimensional space. We denote $\mathbf{0}_n$ the n-dimensional zero vector. From the definition, it is evident that $\bar\be_a\cdot\bh(\bk) = \be_a\cdot\bd_0(\bk)$. The expansion of the rank-1 orthogonal projectors $\hat P_n(\bk) = \frac{1}{4}\big(\hat 1_4+\bb_n(\bk)\cdot\hat \blambda\big)$ involves $\bb_n(\bk)$ given by a closed formula of $\bh(\bk)$; see End Matter. We choose a normalization that is more convenient for the product form, $\text{tr}(\hat\tau_a\hat\sigma_b) = 4$, distinct from those chosen in Ref.~\cite{Graf2021} using generalized Gell-Mann matrices.

The composition of arbitrary matrices $(\bM\cdot\hat \blambda)(\bn\cdot\hat \blambda)$ expanded in $\hat\blambda$ is entirely determined by the fully symmetric and antisymmetric structure constants~\cite{Graf2021} given by $d_{abc} = \frac{1}{8} \text{tr}\big(\{\hat\lambda_a,\hat\lambda_b\}\hat\lambda_c\big)$ and $f_{abc} = -\frac{i}{8}\text{tr}\big([\hat \lambda_a,\hat \lambda_b]\hat \lambda_c\big)$, respectively. The structure constants enable the definition of a star and (generalized) cross product $(\bM\hodge \bn)_a = \sum_{b,c}d_{abc}\,m_b\,n_c$ and $(\bM\times \bn)_a = \sum_{b,c}f_{abc}\, m_b\, n_c$, complementing the scalar product $\bM\cdot\bn = \sum_a m_a\, n_a$~\cite{Graf2021}. Note that the star product vanishes for two-band systems as $\{\hat\sigma_a,\hat\sigma_b\}=2\delta_{ab}$ as well as for Dirac Hamiltonians built solely from $\hat\Gamma_a$. The projector $\hat P_n(\bk)$ involves three terms proportional to $\bh(\bk)$, $\bh_\star(\bk) \equiv \bh(\bk)\hodge\bh(\bk)$, and $\bh_{\star\star}(\bk) \equiv \bh(\bk)\hodge(\bh(\bk)\hodge\bh(\bk))$~\cite{Graf2021}.

We relate the spin-space elements $\bd_0(\bk)$ and $\bd_a(\bk)$ of the Hamiltonian in Eq.~\eqref{eqn:BlochHamiltonian} to the {\it su}(4) algebra by separating $\bh(\bk)$ into its components, e.g., $\bar d_1(\bk) = (d_1(\bk),\mathbf{0}_{14})$ and $\bar \bd_0(\bk) = (\mathbf{0}_3,\bd_0(\bk),\mathbf{0}_9)$. Omitting the momentum dependence for shorter notation, we find the relations~\cite{SupplMat}
\begin{align}
    &\bar\be_a\cdot\big[\bar d_b\hodge\bar \bd_b\big] = d_b\,\bd_b\cdot\be_a \, , \label{eqn:ExampleStarProduct1}\\
    &\bar\be_a\cdot\big[\bar \bd_b\hodge(\bar\bd_b\hodge\bar \bd_0)\big] = \bd_0\cdot\bd_b\,\,\bd_b\cdot\be_a\, ,
    \label{eqn:ExampleStarProduct2}
\end{align}
for $b=1,2,3$ between the star product and the (regular 3-dimensional) scalar product. Furthermore, we obtain
\begin{align}
    &\bar\be_a\cdot\big[\bar d_1\hodge(\bar\bd_2\hodge\bar \bd_3)\big] = -d_1\,\big(\bd_2\times\bd_3\big)\cdot\be_a\, ,
    \label{eqn:ExampleStarProduct3}
\end{align}
relating the star product to the (regular 3-dimensional) cross product between $\bd_2$ and $\bd_3$.

{\it General form of spin polarization.---} Using the explicit form of $\bb_n(\bk)$ and the upper results allows us to calculate the spin polarization of the general spinful two-site Hamiltonian in Eq.~\eqref{eqn:BlochHamiltonian} explicitly via $\langle \hat S_a\rangle_{n}(\bk) = \text{tr}\big[\hat S_a\,\hat P_n(\bk)\big] = \bar\be_a\cdot\bb_n(\bk)$. We decompose the spin polarization into three contributions
\begin{align}
    \langle \hat \bS\rangle_{n}(\bk) = \langle \hat \bS\rangle_{n,0}(\bk) + \langle \hat \bS\rangle_{n,\star}(\bk) + \langle \hat \bS\rangle_{n,\star\star}(\bk) \, ,
    \label{eqn:SpinPolarizationDecomposition}
\end{align}
proportional to $\bar\be_a\cdot\bh$, $\bar\be_a\cdot\bh_\star$, and 
$\bar\be_a\cdot\bh_{\star\star}$, respectively. As outlined in detail in the SM~\cite{SupplMat}, we decompose the Hamiltonian coefficients $\bh$ into their components and identify the nonzero elements, e.g., Eqs.~\eqref{eqn:ExampleStarProduct1}-\eqref{eqn:ExampleStarProduct3}. We give the final general expression in the End Matter and focus on the physical implications in the following. Note that the final expression is independent of the basis choice $\hat \blambda$ during the calculation. For reference, we provide the spin polarization $\langle\hat \bS\rangle_\pm(\bk) = \bd(\bk)/|\bd(\bk)|$ of a spinful (one-site) two-band Hamiltonian $\hat H(\bk) = d_0(\bk)+\bd(\bk)\cdot\bsigma$~\cite{SupplMat}, where the Hamiltonian vector and spin polarization are always aligned. 

\begin{table}
    \centering
    \begin{tabular}{lccc}
        \hline\hline\\[-3mm]
         $\langle \hat \bS\rangle_n(\bk)$ &  \hspace{0mm} Ferromagnet \hspace{0mm} & \hspace{0mm} Altermagnet \hspace{0mm} & \hspace{0mm} Antialtermagnet \hspace{0mm}    \\\hline\hline\\[-3mm]
         $\bM\cdot\bh(\bk)$  & $\bd_0(\bk)$ & $0$   & $0$ \\[1mm]\hline\\[-3mm]
         $\bM\cdot\bh_{\star}(\bk)$ & $0$ &  $d_3(\bk)\,\bd_3(\bk)$  & $0$ \\[1mm]\hline\\[-3mm]
         $\bM\cdot\bh_{\star\star}(\bk)$ & $d_1(\bk)^2\,\bd_0(\bk)$ & $0$   & $d_1(\bk)\,\bd_2(\bk)\!\times\!\bd_3(\bk)$\\\\[-3mm]\hline\hline
    \end{tabular}
    \caption{The three contributions to the spin polarization, sorted by the number of involved {\it su}(4) star products, for characteristic minimal four-band models of ferro-, alter-, and antialtermagnets, enabling a Bloch-state geometric classification.}
    \label{tab:OverviewSpinPolarizationGeneral}
\end{table}

\begin{figure*}
    \centering
    \includegraphics[width=0.75\linewidth]{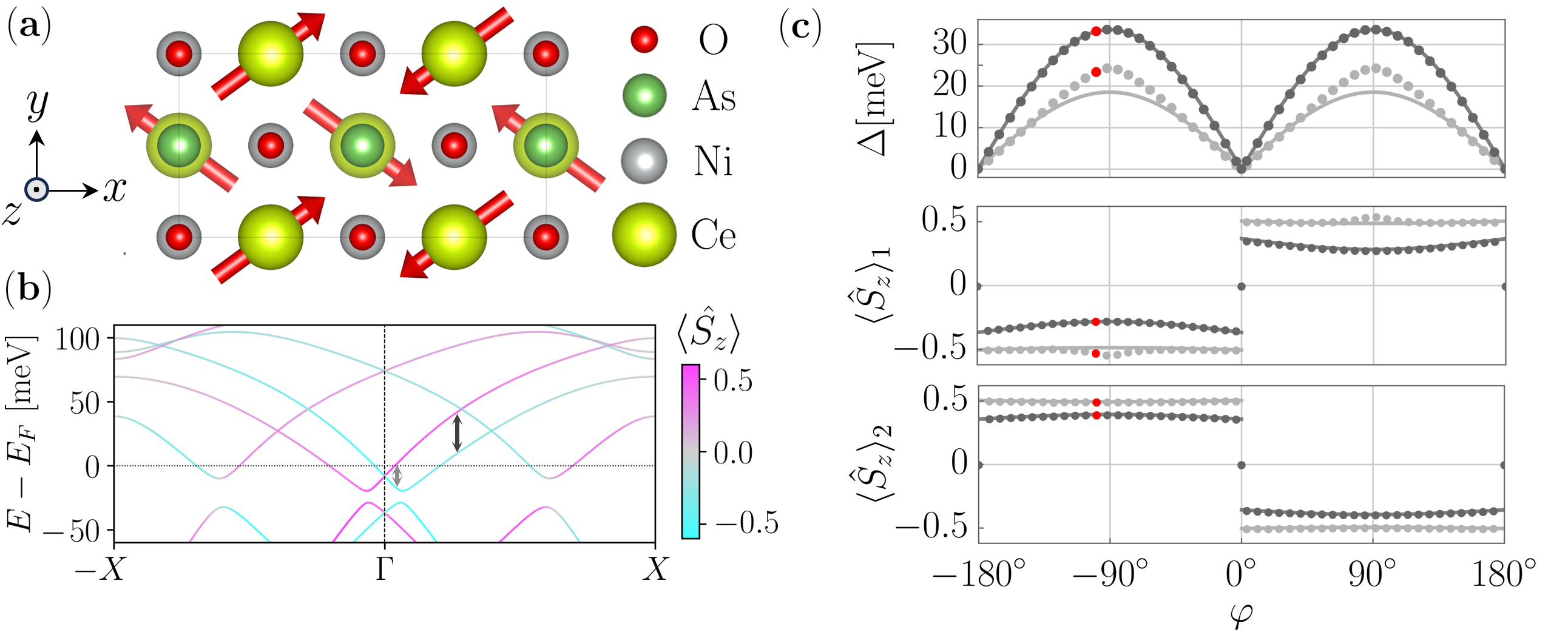}
    \caption{(a) The real-space crystal structure of $\text{CeNiAsO}$ with coplanar non-collinear spin texture. (b) The band structure near the Fermi level with $\Gamma = (0,0,0)$ and $X=(\pi,0,0)$. (c) The angle dependence of spin splitting and spin polarization of the bands 1 and 2 near the Fermi level at two momenta near $\Gamma$, as indicated in (b); the physical angle is marked in red.  }
    \label{fig:DFT}
\end{figure*}

Let us consider a minimal ferromagnetic Hamiltonian of the form $\hat H(\bk) = d_0(\bk)+d_1(\bk)\,\hat\tau_1+\bd_0(\bk)\cdot\hat\bsigma$. As indicated in Tab.~\ref{tab:OverviewSpinPolarizationGeneral}, we obtain two contributions from $\bh$ and $\bh_{\star\star}$, leading to the spin polarization of the lowest band
\begin{align}
    \langle \hat \bS\rangle_1(\bk) \!=\!\frac{1}{E_{1}(\bk)-d_0(\bk)}\!\bigg[\bd_0(\bk)+|d_1(\bk)|\frac{\bd_0(\bk)}{|\bd_0(\bk)|}\bigg] .
\end{align}
The first term is the mean magnetic moment of both sites. We see that the second term corrects the spin polarization induced by hybridization between the two sites, parallel to $\bd_0$. In contrast, the spin polarization of the minimal altermagnetic model $\hat H(\bk) = d_0(\bk)+d_1(\bk)\hat\tau_1+d_3(\bk)\hat\tau_3+\bd_3(\bk)\cdot\hat\tau_3\hat\bsigma$~\cite{Roig2024} is entirely determined by $\bh_\star$, revealing a fundamentally distinct microscopic origin of the unconventional altermagnetic spin splitting. For the lowest band, we obtain
\begin{align}
    \langle \hat \bS\rangle_1(\bk) =  \text{sign}\big(d_3(\bk)\big)\,\frac{\bd_3(\bk)}{|\bd_3(\bk)|} \, ,
\end{align}
showing that the spin polarization aligns with $\bd_3$, interpretable as N\'eel vector for a concrete altermagnetic model, with the sign determined by the $\cP\cT$-breaking site imbalance $d_3$. We note that in both cases, effective two-band descriptions with the same spin polarization are straightforwardly constructed by an adequately chosen $\bd(\bk)$, which physically corresponds to effective spin-dependent hopping due to exchange coupling with the magnetic moments. 

As seen in Eq.~\eqref{eqn:ExampleStarProduct3}, the {\it su}(4) algebra enables a third, fundamentally different form of spin polarization arising in $\bh_{\star\star}$, which we exemplify by the minimal model
$\hat H(\bk) = d_0(\bk)+d_1(\bk)\,\hat\tau_1+\bd_2(\bk)\cdot\hat\tau_2\hat\bsigma+\bd_3(\bk)\cdot\hat\tau_3\hat\bsigma$, which captures $p$-, $f$-, and $h$-wave magnets collectively referred to as antialtermagnets~\cite{Jungwirth2025, jungwirth_symmetry_2026}. For the lowest band, we obtain
\begin{align}
    \langle \hat \bS\rangle_{1}(\bk) = -\frac{d_1(\bk)}{E_{1}(\bk)-d_0(\bk)}\frac{\bd_2(\bk)\times\bd_3(\bk)}{|\bd_2(\bk)\times\bd_3(\bk)|} \, ,
    \label{eqn:SpinPwaveGeneral}
\end{align}
which is perpendicular to the plane spanned by the spin-dependent inter-site and site-imbalance terms $\bd_2$ and $\bd_3$. Inserting the concrete minimal model of a $p$-wave magnet in Eq.~\eqref{eqn:PwaveBlochHamiltonian} confirms the relation to the in-plane magnetic moments in Eq.~\eqref{eqn:OutOfPlanePolarization} and the amplitude for small momenta in Eq.~\eqref{eqn:MomentumDependenceSpinPolarization}; see End Matter for the full expression. We note that out-of-plane spin polarization can also be realized via $d_2\big(\bd_3\times\bd_1\big)$ and $d_3\big(\bd_1\times \bd_2\big)$, as shown in the End Matter, providing guiding principles for realizing antialtermagnets.

\begin{figure}[b!]
    \centering
    \includegraphics[width=0.99\linewidth]{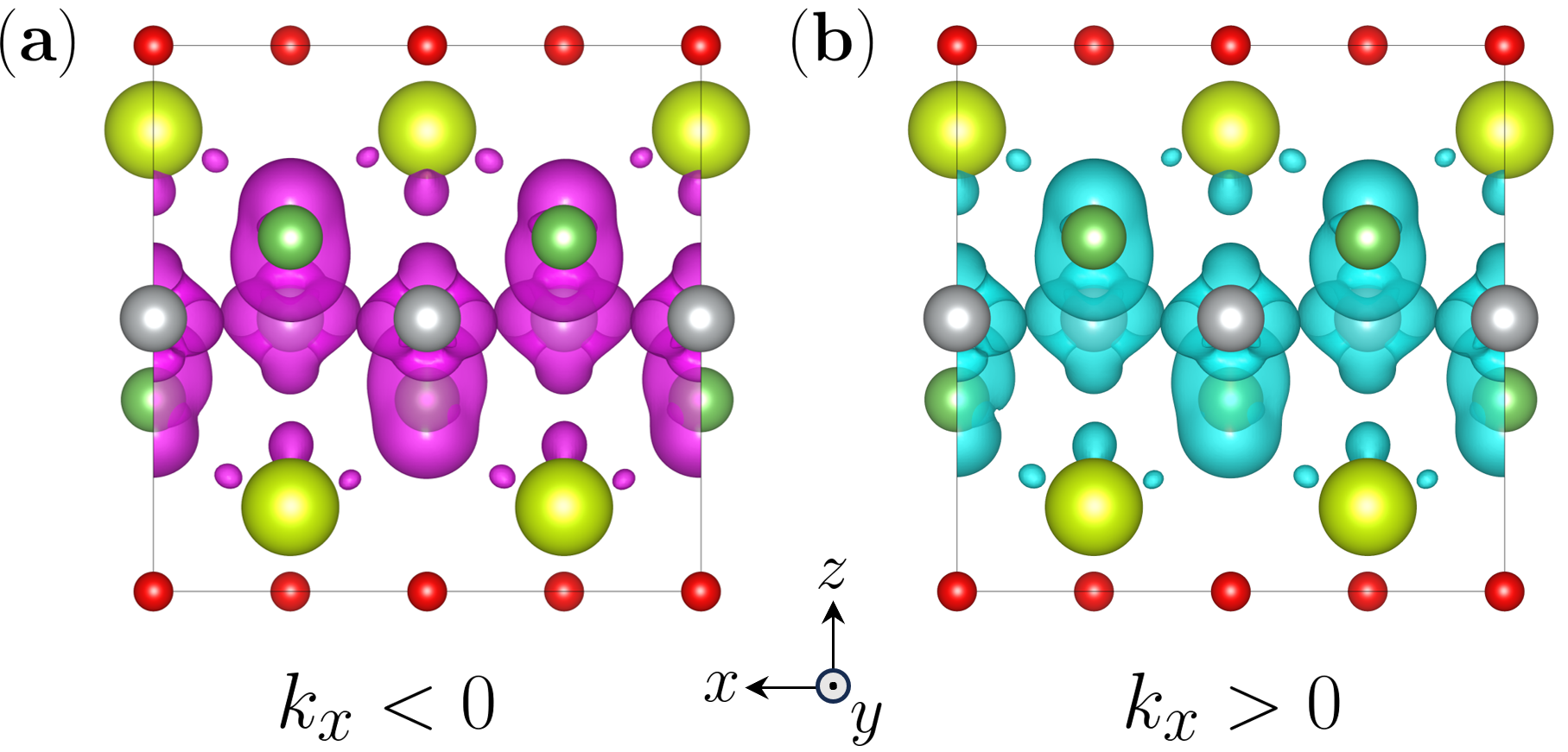}
    \caption{The spin density of band 3 integrated over the left (a) and right (b) half of the Brillouin zone.}
    \label{fig:SpinDensity}
\end{figure}

{\it Application to $\text{CeNiAsO}$.---} In Fig.~\ref{fig:DFT}(a), we show the magnetic unit cell~\cite{Gallego2016, Wu2019} of $\text{CeNiAsO}$, which serves as an antialtermagnetic material candidate~\cite{Hellenes2024, Zhou2025} and provides a motivation for the tight binding model~\cite{Hellenes2024} shown in Fig.~\ref{fig:Model}(a). Here, the magnetic moments of cerium (Ce) atoms are effectively captured via spin-dependent hopping. The characteristic angles of the magnetic moments in the ground state are $\varphi_x \sim 36\degree$, $\varphi_y \sim 144\degree$, resulting in a relative angle $\varphi = -108\degree$. The time-reversal-symmetric spin-splitting is enforced by the combination of spin-translation symmetry $[C_{2z}|| t(\sfrac{1}{2}, 0, 0)]$ and coplanar spin-only symmetry $[C_2\cT||\cT]=[M_z||E]$~\cite{Smejkal2022}, which further enforces the coplanar magnetic order. Additionally, these symmetries imply collinear spin polarization that is perpendicular to the real-space coplanar magnetic order. We show the corresponding band structure with $p$-wave spin splitting and spin polarization along $k_x$ in Fig.~\ref{fig:DFT}(b). The tight-binding Hamiltonian in Eq.~\eqref{eqn:BlochHamiltonian} serves as a minimal model for the four conduction bands at the Fermi level near $\Gamma$. 

To study the ab initio band splitting and spin polarization for varying angles between the in-plane magnetic moments, we artificially constrain the magnetic moments~\cite{SupplMat}. In Fig.~\ref{fig:DFT}(c), we show the splitting energy between the two lowest conduction bands 1 and 2 for two momenta $k_x=0.263\,\pi/a$ (black) and $k_x=0.053\,\pi/a$ (gray) with lattice constant $a$. Near $\Gamma$, we find good overall agreement between the ab initio results (dots) and the predicted angular dependence of the minimal model (solid lines) given in Eqs.~\eqref{eqn:BandGapMainText} and \eqref{eqn:MomentumDependenceSpinPolarization}. Deviations from the minimal four-band prediction are visible near interference with additional bands (gray), where strong hybridization is expected.

We estimate coefficients for $\Delta_\text{eff}(\varphi) = 2c_1|k_x\sin\varphi|$ and $S_{\text{eff},n}(\varphi) = \pm c_{2,n}\,\text{sign}[k_x\sin\varphi]\big[1\mp\frac{1}{2}c_{3,n}|k_x\sin\varphi|\big]$, where the upper and lower signs of $\pm$ and $\mp$ correspond to band 1 and 2, respectively; see solid lines in Fig.~\ref{fig:DFT}(c). We use these coefficients to estimate $t_{n,\text{eff}} = \sqrt{1-c_{3,n}}\,c_1/c_{3,n}$ and $J_{n,\text{eff}}=c_1/\sqrt{c_{3,n}}$. For $k_x=0.263\,\pi/a$ (black), we obtain $t_{\text{eff},1}=24\,\text{meV}$ and $J_{\text{eff},1}=27\,\text{meV}$ for band 1 and $t_{\text{eff},2}=70\,\text{meV}$ and $J_{\text{eff},2}=41\,\text{meV}$ for band 2. Despite the simplicity of the minimal model, the parameters are within the physically reasonable order of magnitude $\sim 10-100\,\text{meV}$ and $J/|t|\lesssim 1$; see Fig.~\ref{fig:DFT}(a). We note that the amplitude is fixed to $c_{2,\pm}=\sqrt{1-c_{3,\pm}}$ and overestimates the amplitude of $\langle \hat S_z\rangle_n$ by a factor of two. Better quantitative agreement requires an extended tight-binding description capturing the band structure over the full Brillouin zone (BZ).

Our ab initio calculation enables us to approach the real-space distribution of the spin polarization for individual Kohn-Sham orbitals in the form of $\langle \hat S_z\rangle_n(\br,\bk)$. While the overall spin polarization, integrated over the full BZ, is locally-compensated $\int_\text{BZ}\langle \hat S_z\rangle_n(\br,\bk) = 0$, the momentum-resolved spin polarization is nonzero; see Fig.~\ref{fig:SpinDensity}(a,b). To give an estimate of the amplitude, we find that the real-space-integrated momentum-projected spin density is 0.24 with respect to the corresponding projected charge density. Our identification of this "hidden" alternating intra-site spin order adds yet another meaning to the term antialtermagnets. So far, the $p$-, $f$-, or $h$-wave magnets are jointly referred to as antialtermagnets due to the combination of alternating spin polarization in momentum space, akin to altermagnets, with antiferroic direct space order of sublattice magnetic dipoles~\cite{Jungwirth2025, jungwirth_symmetry_2026}. In addition, we note that antialtermagnetism arises from the antiparallel orientation of site-spin densities in direct space that compensate each other, but emerge when projected over the $k_x<0$ or $k_x>0$ half of the BZ; see Fig.~\ref{fig:Model} and \ref{fig:SpinDensity}(a,b).

{\it Conclusion and Outlook.---} We have employed the analytic band projector expressions in terms of {\it su}(4) star products of the Hamiltonian vector and band dispersions~\cite{Graf2021} to link the geometric structure of the Bloch Hamiltonian to the real-space non-collinear magnetic texture. We have shown that the out-of-plane spin polarization predicted for $p$-wave magnets~\cite{Hellenes2024} is proportional to the cross product of the in-plane magnetic moments, supported by ab initio results of \text{CeNiAsO}.

By providing the full analytic expression for the spin polarization of all spinful two-site Hamiltonians, we have obtained a Bloch-state geometric classification, contrasting ferro-, alter-, and antialtermagnetism; see Tab.~\ref{tab:OverviewSpinPolarizationGeneral}. The geometric classification through the star product of the {\it su}(4) algebra suggests that antialtermagnets, and potentially other non-collinear magnets, yield platforms for nontrivial quantum geometry, arising from the interplay between the star product and nontrivial momentum dependence~\cite{Barnett2012, Graf2021, Kemp2022, Mercaldo2023, Mitscherling2025a}, for which at least a four-band description is required.

Our general result supports the intuitive understanding that $p$-wave magnetism arises from the chirality of the coplanar magnetic ordering that is felt by electrons propagating parallel to the direction implied by the defining combined time-reversal-translation $\cT t$ symmetry of $p$-wave magnets, as sketched in Fig.~\ref{fig:Model}(c). Pictorially, the non-collinear order leads to electrons rotating in spin space at nonzero momentum, such that the in-plane spin component averages over a single unit cell, while it induces a nonzero out-of-plane component. From the real-space perspective, our ab initio results have revealed a corresponding antiparallel intra-site spin order; see Fig.~\ref{fig:Model}(d), complementing the real- and momentum-space perspective of $p$-wave magnets.

{\it Acknowledgment.---} We thank Dan S. Borgnia and Kush Saha for valuable discussions. J.M. was supported, in part, by the Deutsche Forschungsgemeinschaft under Grant cluster of excellence ct.qmat (EXC 2147, Project No. 390858490). L\v{S} acknowledges funding from the ERC Starting Grant No. 101165122 and Deutsche Forschungsgemeinschaft (DFG) grant no. TRR 288 - 7422213477 (Projects A09 and B05). This work was supported by the Ministry of Education, Youth and Sports of the Czech Republic through the e-INFRA CZ (ID:90254).

\bibliography{biblio}

\section*{End matter}

\subsection{Rank-1 projectors of four-band systems}

We adjust the result by Graf and Pi\'echon~\cite{Graf2021} to expressions best suitable for the basis built upon the Kronecker product of site- and spin-space Pauli matrices $\hat\tau_a\hat\sigma_b$ and the corresponding normalization $\text{tr}(\hat\tau_a\hat\sigma_b)=4$. We define the expansion of the rank-1 projection operator, slightly distinct from Ref.~\cite{Graf2021}, as
\begin{align}
    \hat P_n(\bk) = \frac{1}{4}\Big(\hat 1_4+\bb_n(\bk)\cdot\hat \blambda\Big) \, .
\end{align}
and omit the momentum dependence for shorter notation in the following. The Bloch vector takes the form
\begin{align}
    \bb_n = c_{n,0}\,\bh +  c_{n,\star}\,\bh_\star +  c_{n,\star\star}\,\bh_{\star\star} \, .
    \label{eqn:BlochVectorEM}
\end{align}
with star products $\bh_\star = \bh\hodge\bh$ and $\bh_{\star\star}=\bh\star(\bh\star\bh)$ of Hamiltonian vector $h_a(\bk) = \frac{1}{4}\text{tr}\big[\hat H(\bk)\,\hat \lambda_a\big]$. The prefactors in Eq.~\eqref{eqn:BlochVectorEM} read
\begin{align}
    &c_{n,0} = \frac{\big(E_n-d_0\big)^2-|\bh|^2}{\big(E_n-d_0\big)^3-\big(E_n-d_0\big)\,|\bh|^2-\frac{1}{3}\bh\cdot \bh_\star} \,, \label{eqn:prefactor1}\\[2mm]
    &c_{n,\star} = \frac{E_n-d_0}{\big(E_n-d_0\big)^3-\big(E_n-d_0\big)\,|\bh|^2-\frac{1}{3}\bh\cdot \bh_\star} \, , \label{eqn:prefactor2}\\[2mm]
    &c_{n,\star\star} = \frac{1}{\big(E_n-d_0\big)^3-\big(E_n-d_0\big)\,|\bh|^2-\frac{1}{3}\bh\cdot \bh_\star} \, , \label{eqn:prefactor3}
\end{align}
with $d_0(\bk) = \frac{1}{4}\text{tr}\,\hat H(\bk)$. 

\subsection{Spin polarization of general spinful two-site Hamiltonian in Eq.~\eqref{eqn:BlochHamiltonian}}

The spin polarization $\langle \hat \bS\rangle_n(\bk) = \big(\langle \hat S_a\rangle_n(\bk)\big)$ of band $n$ in Eq.~\eqref{eqn:SpinPolarizationDecomposition} yields the three contributions
\begin{align}
    &\langle \hat S_a\rangle_{n,0}(\bk) = c_{n,0}(\bk)\,\bar\be_a\cdot\bh(\bk) \, , \label{eqn:SpinGeneral1}\\ &\langle \hat S_a\rangle_{n,\star}(\bk) =c_{n,\star}(\bk)\,\bar\be_a\cdot\bh_\star(\bk)\, , \label{eqn:SpinGeneral2}\\&\langle \hat S_a\rangle_{n,\star\star}(\bk) =c_{n,\star\star}(\bk)\,\bar\be_a\cdot\bh_{\star\star}(\bk)\, , \label{eqn:SpinGeneral3}
\end{align}
involving the coefficients of the Bloch Hamiltonian $\bh(\bk)$ and the spin operator $\bar\be_a$ in the orthogonal matrix basis $\hat\blambda$. The coefficients are given in Eqs.~\eqref{eqn:prefactor1}-\eqref{eqn:prefactor3}. We omit the momentum dependence for shorter notation in the following. From the definition, it is evident that 
\begin{align}
    \bar\be_a\cdot\bh &= \be_a\cdot\bd_0 \, ,
    \label{eqn:SpinPart1} 
\end{align}
such that $\langle \hat \bS\rangle_{n,0}\propto \bd_0$, which closely resembles the generic results for spinful one-site (two-band) Hamiltonians $\hat H = \bd_0\cdot\bsigma$. In contrast, the four-band structure enables a nontrivial structure arising from the interplay between site- and spin-space. The second contribution~\eqref{eqn:SpinGeneral2} reads
\begin{align}
    \bar\be_a\cdot\bh_\star &= 2\,\be_a\cdot\big[d_1\,\bd_1+d_2\,\bd_2+d_3\,\bd_3\big] \, ,
    \label{eqn:SpinPart2}
\end{align}
enabling spin polarization as a superposition of the spin-dependent couplings $\bd_a$ weighted by the corresponding spin-independent site couplings $d_a$. The third contribution~\eqref{eqn:SpinGeneral3} yields three distinct terms,
\begin{align}
    &\bar\be_a\cdot\bh_{\star\star} =  2\,\be_a\cdot\big[(d_1\,d_1+d_2\,d_2+d_3\,d_3)\,\bd_0 \nonumber \\[1mm]& \,\,\,\,\,\,\,\,   +(\bd_1\cdot\bd_0)\,\bd_1+(\bd_2\cdot\bd_0)\,\bd_2+(\bd_3\cdot\bd_0)\,\bd_3 \nonumber \\[1mm]    &\,\,\,\,\,\,\,\,-d_1\big(\bd_2\times\bd_3\big)-d_2\big(\bd_3\times\bd_1\big)-d_3\big(\bd_1\times \bd_2\big)\big] \, ,
    \label{eqn:SpinPart3}
\end{align}
where the first line can be interpreted as correction to Eq.~\eqref{eqn:SpinPart1} and the second line captures the interplay of $\bd_0$ and $\bd_a$. The third line reveals the possibility of spin polarization orthogonal to coplanar magnetic moments, as exemplified by the minimal $p$-wave model in Eq.~\eqref{eqn:PwaveBlochHamiltonian}. 

\begin{widetext}
\subsection{Analytic expression for minimal $p$-wave model}

We give the full analytic expressions for the minimal $p$-wave model in Eq.~\eqref{eqn:PwaveBlochHamiltonian}. The four band dispersions read
\begin{align}
    &E_{(ns)}(\bk) = 2t\cos k_y + 2|t|\,n\,\sqrt{\cos^2(k_x/2)+\Big(\frac{J}{t}\Big)^2\Big[\sin^2 (k_x/2)+\cos^2(k_y)+2\,s\,|\sin(k_x/2)\cos (k_y)\sin\varphi |\Big]}\, .
\end{align}
where $n,s=\pm$. The lowest two bands are $E_1(\bk) = E_{(-+)}(\bk)$ and $E_2(\bk) = E_{(--)}(\bk)$. The corresponding spin polarization reads
\begin{align}
    \langle \hat\bS\rangle_{(ns)}(\bk)&=-n\,s\frac{\text{sign}[t\,\sin(k_x/2)\,\cos k_y\,\sin\varphi]\cos(k_x/2)}{\sqrt{\cos^2(k_x/2)+\Big(\frac{J}{t}\Big)^2\Big[\sin^2 (k_x/2)+\cos^2(k_y)+2\,s\,|\sin(k_x/2)\cos (k_y)\sin\varphi |\Big]}}\,\be_z \, .
    \label{eqn:PolarizationAmplitude}
\end{align}

\clearpage

\end{widetext}


\appendix

\newpage
\clearpage

\begin{widetext}

\newpage

\section*{Supplemental Material for \\[2mm] ``Microscopic origin of $p$-wave magnetism''}

In the Supplemental Material (SM), we provide (A) the band structure and spin polarization of the two-band model for reference. We discuss (B) several basis choices for four-band Bloch Hamiltonians in terms of representation of the {\it su}(4) algebra and Dirac matrices, which we use to derive (C) the spin polarization of a general spinful two-site tight-binding Hamiltonian, which we exemplify for various minimal Hamiltonians. We apply the general theory to (D) the tight-binding model for p-wave magnets. We conclude by (E) summarizing the ab initio density-functional calculations on \text{CeNiAsO}.

\section{Band structure and spin polarization of two-band model}

We introduce the basic notation and general results for the tight-binding models in the following.

\subsection{General two-band Hamiltonian}

We consider a lattice system in three spatial dimensions characterized within a right-handed orthonormal coordinate system $\be_x$, $\be_y$, and $\be_z$ with a single site A per unit cell in addition to the two spin-$\frac{1}{2}$ degrees of freedom labeled as $\uparrow$ and $\downarrow$ with respect to the quantization axis $\be_z$ aligned with the spatial coordinate system. The lattice-translational and quadratic Hamiltonian generally reads 
\begin{align}
    \hat H = \sum_\bk \hat c^\dagger_\bk \,\hat H(\bk)\,\hat c^{}_\bk \, ,
    \label{BlochHamiltonian2Band}
\end{align}
where we denote the creation operators of the Bloch electrons as $\hat c^\dagger_\bk = \big(\hat c^\dagger_{\uparrow,\bk},\, \hat c^\dagger_{\downarrow,\bk}\big)$ and the annihilation operators $\hat c^{}_\bk$, analogously. The spin operators within this basis take the form
\begin{align}
    \hat S_x = \hat \sigma_1 = \begin{pmatrix}
        0 & 1 \\ 1 & 0
    \end{pmatrix}\, , \hspace{1cm}
    \hat S_y = \hat \sigma_2 = \begin{pmatrix}
        0 & -i \\ i & 0
    \end{pmatrix}\, , \hspace{1cm}
    \hat S_z = \hat \sigma_3 = \begin{pmatrix}
        1 & 0 \\ 0 & -1
    \end{pmatrix}\, .
\end{align}
in units $\hbar/2$, involving the Pauli matrices $\hat \sigma_a$. We expand the Bloch Hamiltonian matrix $\hat H(\bk)$ in the Pauli matrices, defining 
\begin{align}
    \hat H(\bk) = d_0(\bk)\,\hat 1+\bd(\bk)\cdot\hat \bsigma \, ,
\end{align}
where $\hat \bsigma = \big(\hat \sigma_1,\hat \sigma_2,\hat \sigma_3\big)$. We denote the unit matrix $\hat 1 = \hat \sigma_0$ synonymously. The Pauli matrices are traceless, leading to $\text{tr}\,\hat H(\bk) = 2d_0(\bk)$. The identity and Pauli matrices form an orthonormal basis with respect to the Hilbert-Schmidt inner product
\begin{align}
    \langle \hat A,\hat B\rangle = \text{tr}\big[\hat A^\dagger\hat B\big].
\end{align}
up to a constant rescaling, satisfying $\langle \hat \sigma_a,\hat \sigma_b\rangle = 2\delta_{ab}$. The band dispersions of the two bands read 
\begin{align}
    E_\pm(\bk) = d_0(\bk)\pm |\bd(\bk)|
\end{align}
with corresponding orthogonal band projectors
\begin{align}
    \hat P_\pm(\bk) = \frac{1}{2}\Big(\hat 1\pm\bn(\bk)\cdot\hat \bsigma\Big)
\end{align}
with $\bn(\bk) = \big(n_1(\bk),n_2(\bk),n_3(\bk)\big) = \bd(\bk)/|\bd(\bk)|$, satisfying $\hat P_n(\bk)\hat P_m(\bk) = \delta_{nm}\hat P_n(\bk)$ and $\hat H(\bk)\hat P_n(\bk) = E_n(\bk)\hat P_n(\bk)$. Via this form, it is straightforward to show that the spin polarization distribution over the Brillouin zone is captured by the Bloch vector $\bn(\bk)$ via
\begin{align}
    &\langle \hat S_x\rangle_{\pm}(\bk) \equiv \text{tr}\big[\hat S_x\,\hat P_\pm(\bk)\big] = \pm n_1(\bk)  \, ,\\
    &\langle \hat S_y\rangle_{\pm}(\bk) \equiv \text{tr}\big[\hat S_y\,\hat P_\pm(\bk)\big] = \pm n_2(\bk) \, ,\\
    &\langle \hat S_z\rangle_{\pm}(\bk) \equiv \text{tr}\big[\hat S_z\,\hat P_\pm(\bk)\big] = \pm n_3(\bk) \, .
\end{align}
or in compact vector notation 
\begin{align}
    \langle \hat \bS\rangle_{\pm}(\bk) = \pm \bn(\bk) \, .
    \label{eqn:SpinPolarization2bandModel}
\end{align}
The time-reversal operator reads 
\begin{align}
    \hat \cT = i\hat \sigma_2\,\hat \cK
\end{align}
with complex conjugation operator $\hat \cK$. The operator is antiunitary, satisfying $\hat \cT^2 = -\hat \sigma_2\hat \sigma_2 = -\hat 1$ using that $\sigma_a^2=\hat 1$ and $i\hat \sigma_2$ is real. It reverses the spin operator since 
\begin{align}
    &\hat \cT\,\hat S_x\,\hat \cT^{-1} = i\hat \sigma_2\hat \cK \hat \sigma_1(-\hat \cK i\hat \sigma_2) = \hat \sigma_2\hat \sigma_1\hat \sigma_2 = -\hat \sigma_1 = -\hat S_x \, , \label{eqn:actionTx}\\
    &\hat \cT\,\hat S_y\,\hat \cT^{-1} = i\hat \sigma_2\hat \cK \hat \sigma_2(-\hat \cK i\hat \sigma_2) = -\hat \sigma_2\hat \sigma_2\hat \sigma_2 = -\hat \sigma_2 = -\hat S_y \, , \label{eqn:actionTy}\\
    &\hat \cT\,\hat S_z\,\hat \cT^{-1} = i\hat \sigma_2\hat \cK \hat \sigma_3(-\hat \cK i\hat \sigma_2) = \hat \sigma_2\hat \sigma_3\hat \sigma_2 = -\hat \sigma_3 = -\hat S_z \, . \label{eqn:actionTz}
\end{align}

\subsection{Minimal model with magnetic exchange fields}

We consider a square lattice in the x-y plane spanned by unit vectors $\be_x$ and $\be_y$ with real spin-independent hopping amplitude $t$. Furthermore, we include spin-dependent hopping in x and y directions over local magnetic moments $\bJ_x$ and $\bJ_y$, respectively, complemented by a local magnetic moment $\bJ_0$ at the lattice site. The corresponding tight-binding Hamiltonian reads
\begin{align}
    \hat H = t\sum_{\bR,\sigma}\,\big(\hat c^\dagger_{\bR+\be_x}\hat c^{}_{\bR}+\hat c^\dagger_{\bR+\be_y}\hat c^{}_{\bR}+\text{h.c.}\big) + \sum_\bR\big(\hat c^\dagger_{\bR+\be_x}\,\bJ_x\cdot\hat\bsigma\,\hat c^{}_\bR+\hat c^\dagger_{\bR+\be_y}\,\bJ_y\cdot\hat\bsigma\,\hat c^{}_\bR+\text{h.c.}\big) + \sum_\bR\,\hat c^\dagger_{\bR}\,\bJ_0\cdot\hat\bsigma\,\hat c^{}_\bR \, .
\end{align}
with h.c.~denoting complex conjugation. We obtain the Bloch Hamiltonian \eqref{BlochHamiltonian2Band} after Fourier transformation with 
\begin{align}
    \hat c^\dagger_{\bR,\nu}=\frac{1}{\sqrt{N_c}}\sum_\bk e^{-i\bk\cdot(\bR+\br_\alpha)}\,\hat c^\dagger_{\bk,\nu}
\end{align}
setting the single lattice site $\nu=A$ to the unit cell origin $\br_A=0$ for convenience. $N_c$ is the number of unit cells. The Bloch Hamiltonian reads
\begin{align}
    \hat H(\bk) = 2t\big(\cos k_x+\cos k_y\big)\,\hat 1_2+ \big(2\cos k_x\,\bJ_x+2\cos k_y\,\bJ_y+\bJ_0\big)\cdot\hat\bsigma \, .
\end{align}
Setting $\bJ_x = J\cos\varphi\, \be_x+J\sin\varphi\,\be_y$ with $\bJ_y=-\bJ_x$ and $\bJ_0=0$ serves as a minimal altermagnetic model with in-plane magnetic moments~\cite{Smejkal2022c}. We obtain
\begin{align}
    \bd(\bk) = 2J\big(\cos k_x-\cos k_y\big)\big(\cos\varphi\, \be_x+\sin\varphi\,\be_y\big)
\end{align}
with $|\bd(\bk)| = 2|J||\cos k_x-\cos k_y|$ leading to the band dispersion 
\begin{align}
    E_\pm(\bk) = 2t\big(\cos k_x+\cos k_y\big)\pm 2|J||\cos k_x-\cos k_y|
\end{align}
and spin polarization
\begin{align}
    \langle \hat \bS\rangle_{\pm}(\bk) = \pm \text{sign}\,J\,\text{sign}(\cos k_x-\cos k_y)\,\big(\cos\varphi\, \be_x+\sin\varphi\,\be_y\big) \, .
\end{align}
We see that the spin polarization aligns with the magnetic moment, as expected by the general discussion in Eq.~\eqref{eqn:SpinPolarization2bandModel}. 

\section{Basis choices for four-band Bloch Hamiltonians}

We extend the analysis to two sites per unit cell in addition to the spin degree of freedom, resulting in a four-band Bloch Hamiltonian
\begin{align}
    \hat H = \sum_\bk \hat c^\dagger_\bk \,\hat H(\bk)\,\hat c^{}_\bk\,
\end{align}
with the basis choice $\hat c^\dagger_\bk=\big(\hat c^\dagger_{A,\uparrow}(\bk),\hat c^\dagger_{A,\downarrow}(\bk),\hat c^\dagger_{B,\uparrow}(\bk),\hat c^\dagger_{B,\downarrow}(\bk)\big)$. The Hermitian $4\times4$ Bloch Hamiltonian $\hat H(\bk)$ has 16 independent real components. 

\subsection{Motivating the form given in Eq.~\eqref{eqn:BlochHamiltonian}}

In the following, we employ the physically meaningful decomposition into spin and site components, harnessing the underlying block structure of the Bloch Hamiltonian in the site-spin basis. For this, we express the Bloch Hamiltonian $\hat H(\bk)$ in terms of Pauli matrices $\hat \tau_a$ and $\hat \sigma_b$ for the site and spin degree of freedom, respectively, and the identity matrix $\hat \tau_0=\hat \sigma_0=\hat 1$, expanding
\begin{align}
    \hat H(\bk) = \sum_{a,b}h_{ab}(\bk)\,\hat \tau_a\otimes \hat \sigma_b \, .
\end{align}
in terms of the 16 basis elements $\hat \tau_a\otimes \hat \sigma_b$ with Kronecker product $\otimes$. The basis is orthonormal with respect to the Hilbert-Schmidt inner product
\begin{align}
    \langle \hat \tau_a\otimes\hat \sigma_b,\hat \tau_{a'}\otimes\hat \sigma_{b'}\rangle = \text{tr}\big[(\hat \tau_a\otimes\hat \sigma_b)(\hat \tau_{a'}\otimes\hat \sigma_{b'})\big] = \text{tr}\big[\hat \tau_a\hat\tau_{a'}\big]\text{tr}\big[\hat \sigma_b\hat \sigma_{b'}\big] = 4\,\delta_{aa'}\,\delta_{bb'}.
    \label{eqn:normalization}
\end{align}
up to a constant rescaling. Note that the normalization for the two- and four-band cases differs. We have $\text{tr}[\hat \tau_a\otimes \hat \sigma_b]=0$ for $a=b\neq 0$. The components of the spin operator in units $\hbar/2$ take the form 
\begin{align}
    \hat S_x = \hat 1 \otimes \hat\sigma_1 = \begin{pmatrix} \hat\sigma_1 & 0 \\ 0 & \hat\sigma_1 \end{pmatrix} \, , \hspace{1cm} \hat S_y = \hat 1 \otimes \hat\sigma_2 = \begin{pmatrix} \hat\sigma_2 & 0 \\ 0 & \hat\sigma_2 \end{pmatrix} \, , \hspace{1cm} 
    \hat S_z = \hat 1 \otimes \hat\sigma_3 = \begin{pmatrix} \hat\sigma_3 & 0 \\ 0 & \hat\sigma_3 \end{pmatrix} \, ,
\end{align}
where we align the spin quantization axes with the spatial right-handed coordinate system $\be_x$, $\be_y$, and $\be_z$. In light of the spin-dependent hopping amplitudes, it is convenient to combine the spin Pauli matrices as $\hat\bsigma=(\hat \sigma_1,\hat\sigma_2,\hat \sigma_3)$ and separate the $4\times 4$ Bloch Hamiltonian matrix in the $2\times 2$ block structure
\begin{align}
    \hat H(\bk) = \begin{pmatrix}
    \hat H_A(\bk) & \hat H_{AB}(\bk) \\ \hat H_{AB}^\dagger(\bk) & \hat H_B(\bk)
    \end{pmatrix} \, ,
\end{align}
where the $2\times 2$ matrices $\hat H_A(\bk)$ and $\hat H_B(\bk)$ captures the on-site energies of and hopping between the $A$ and $B$ sublattices. As these matrices are Hermitian, we can express them as 
\begin{align}
    \hat H_\nu(\bk) = d_{\nu}(\bk)+\bd_\nu(\bk)\cdot\hat\bsigma \, ,
    \label{eqn:expansionHalpha}
\end{align}
where the scalar $d_{\nu}(\bk)$ captures the onsite energy and spin-independent hopping. We omit $\hat 1$ for shorter notation, if unambiguous. The (three-)vectors $\bd_\nu(\bk)$ yield on-site Zeeman splitting as well as (intra-sublattice) spin-dependent hopping. In Eq.~\eqref{eqn:BlochHamiltonian}, we decomposed the corresponding part of the Hamiltonian in terms of $\hat 1\equiv\hat\tau_0\otimes \hat\sigma_0$, $\hat\bsigma \equiv \hat\tau_0\otimes \hat\bsigma$, $\hat \tau_3\equiv\hat\tau_3\otimes \hat\sigma_0$, and $\hat\tau_3\otimes\hat\bsigma$, leading to the relation
\begin{align}
    &d_0(\bk) = \frac{1}{2}\big(d_{A}(\bk)+d_{B}(\bk)\big) \, ,\\
    &d_3(\bk) = \frac{1}{2}\big(d_{A}(\bk)-d_{B}(\bk)\big) \, ,\\
    &\bd_0(\bk) = \frac{1}{2}\big(\bd_{A}(\bk)+\bd_{B}(\bk)\big) \, ,\\
    &\bd_3(\bk) = \frac{1}{2}\big(\bd_{A}(\bk)-\bd_{B}(\bk)\big) \, .
\end{align}
The inter-site hopping between $A$ and $B$ are captured by the (in general, non-Hermitian) $2\times 2$ matrix $\hat H_{AB}(\bk)$. In Eq.~\eqref{eqn:BlochHamiltonian}, we expanded the eight independent components of $\hat H_{AB}(\bk)$ in terms of the spin-independent hopping $d_1(\bk)\,\hat\tau_1$ and $d_2(\bk)\,\hat\tau_2$ and spin-dependent hopping $\bd_1(\bk)\cdot\hat\tau_1\hat\bsigma\equiv \hat\tau_1\otimes \bd_1(\bk)\cdot\hat\bsigma$ and $\bd_2(\bk)\cdot\hat\tau_2\hat\bsigma\equiv \hat\tau_2\otimes \bd_2(\bk)\cdot\hat\bsigma$.

\subsection{Representation of {\it su}(4) algebra}

We use algebraic structure of the 15 traceless $4\times 4$ Hermitian matrices $\hat \tau_a\otimes \hat \sigma_b$, to which we refer to as $\hat \lambda_a$ and combine into $\hat \blambda = (\hat \lambda_1, ..., \hat\lambda_{15})$ in the following. We can expand the product of two $\hat \lambda_a$ as \cite{Kaplan1967}
\begin{align}
    \hat \lambda_a\hat \lambda_b = \delta_{ab}\,\hat 1_4+d_{abc}\,\hat \lambda_c+i\,f_{abc}\,\hat \lambda_c
    \label{eqn:productRelation}
\end{align}
with real structure constants $d_{abc}$  and $f_{abc}$. We imply summation over equal indices. Using Eq.~\eqref{eqn:normalization}, it follows
\begin{align}
    &\big[\hat \lambda_a,\hat \lambda_b \big] = 2\,i\,f_{abc}\,\hat \lambda_c \, , \\
    &\big\{\hat \lambda_a,\hat \lambda_b\big\} = 2\,\delta_{ab}\,\hat 1_4 + 2\,d_{abc}\,\hat \lambda_c \, , \\
    &\text{tr}\big(\hat\lambda_a\hat\lambda_b\big) = 4\,\delta_{ab} \, ,
    \label{eqn:normalizationGamma}
\end{align}
relating $f_{abc}$ and $d_{abc}$ to the commutator and anticommutator of the $\hat \lambda_a$. Note that the (generalized) Gell-Mann matrices $\hat \lambda^{GM}_a$ offer another basis for $4\times 4$ traceless Hermitian matrices with a different normalization $\text{tr}(\hat \lambda^{GM}_a \hat \lambda^{GM}_b) = 2\delta_{ab}$ \cite{GellMann1962, Bertlmann2008, Graf2021}. The Gell-Mann matrices do not reflect the product structure arising from two sites combined with two spins and, thus, yield a less convenient basis choice. The structure constants are obtained by
\begin{align}
    &f_{abc} = -\frac{i}{8}\text{tr}\big([\hat \lambda_a,\hat \lambda_b]\hat \lambda_c\big) \\
    &d_{abc} = \frac{1}{8} \text{tr}\big(\{\hat\lambda_a,\hat\lambda_b\}\hat\lambda_c\big) \, ,
\end{align}
which are fully antisymmetric and symmetric under all indices exchanges, respectively. The relation~\eqref{eqn:productRelation} for general Hermitian $4\times 4$ matrices $\bM\cdot\hat \blambda$ and $\bn\cdot\hat \blambda$ expanded within a fixed basis choice takes the compact form
\begin{align}
    \big(\bM\cdot\hat \blambda\big)\big(\bn\cdot\hat \blambda\big) = \bM\cdot\bn\,\hat 1_4+(\bM\hodge\bn+i\,\bM\times \bn)\cdot\hat \blambda \, ,
    \label{eqn:productFormula}
\end{align}
where we defined the scalar product, star product, and cross product via \cite{Graf2021}
\begin{align}
    &\bM\cdot\bn = m_a n_a \, , \label{eqn:scalarSU}\\
    &(\bM\hodge \bn)_a = d_{abc} m_b n_c  \, , \label{eqn:starSU}\\
    &(\bM\times \bn)_a = f_{abc} m_b n_c  \, .  \label{eqn:crossSU}
\end{align}
In the following, we expand the Bloch Hamiltonian as 
\begin{align}
    \hat H(\bk) = d_0(\bk)\hat 1_4+\bh(\bk)\cdot\hat \blambda
\end{align}
where $d_0(\bk) = \frac{1}{4}\text{tr}\,\hat H(\bk)$ and $h_a(\bk) = \frac{1}{4}\text{tr}\big[\hat H(\bk)\,\hat \lambda_a\big]$ with $\bh(\bk) = \big(h_1(\bk), ..., h_{15}(\bk)\big)$. We focus on Hamiltonians with band dispersions $E_n(\bk)$ that are degenerate at most at isolated momenta and define the composition of the corresponding orthonormal band projectors as
\begin{align}
    \hat P_n(\bk) = \frac{1}{4}\Big(\hat 1_4+\bb_n(\bk)\cdot\hat \blambda\Big) \, ,
    \label{eqn:projectorBvector}
\end{align}
satisfying $\text{tr}\,\hat P_n(\bk) = 1$ with $\bb_n(\bk) = \text{tr}\big[\hat P_n(\bk)\,\hat \blambda\big]$. Note the rescaling of $\bb_n(\bk)$ with respect to Ref.~\cite{Graf2021}, which we choose due to the distinct normalization of the {\it su}(4) basis in Eq.~\eqref{eqn:normalization}. We re-derive the expression for $\bb_n(\bk)$ arising from the band projector in terms of the band dispersion and Bloch Hamiltonian $\hat P_n\big(E_n(\bk), \hat H(\bk)\big)$ \cite{Graf2021} for the different normalization suitable for decompositions of four-band Bloch Hamiltonians expressed in terms of $\hat\tau_a\otimes \hat\sigma_b$. The relevant Casimir invariants read
\begin{align}
    &C_2 = \text{tr}\big[\big(\hat H(\bk)-d_0(\bk)\hat 1_4\big)^2\big] = \text{tr}\big[\big(\bh(\bk)\cdot\hat \blambda\big)\big(\bh(\bk)\cdot\hat \blambda\big)\big] = 4\,\bh(\bk)\cdot\bh(\bk) = 4|\bh(\bk)|^2 \, , \\
     &C_3 = \text{tr}\big[\big(\hat H(\bk)-d_0(\bk)\hat 1_4\big)^3\big] = \text{tr}\big[\big(\bh(\bk)\cdot\hat \blambda\big)\big(\bh(\bk)\cdot\hat \blambda\big)\big(\bh(\bk)\cdot\hat \blambda\big)\big] = 4\,\bh(\bk)\cdot\bh_\star(\bk)\, ,
\end{align}
with $\bh_\star(\bk) = \bh(\bk)\hodge\bh(\bk)$~\cite{Graf2021}. The band projector involves the traceless part of the Bloch Hamiltonian up to the third power, such that we calculate
\begin{align}
    &\text{tr}\big[\big(\bh(\bk)\cdot\hat \blambda\big)\,\hat \blambda\big] = 4\,\bh(\bk) \, ,\\
    &\text{tr}\big[\big(\bh(\bk)\cdot\hat \blambda\big)^2\,\hat \blambda\big] = 4\,\bh_\star(\bk) \, , \\
    &\text{tr}\big[\big(\bh(\bk)\cdot\hat \blambda\big)^3\,\hat \blambda\big] = 4\,|\bh(\bk)|^2+4\,\bh_{\star\star}(\bk) \, ,
\end{align}
using $\big(\bh(\bk)\hodge\bh(\bk)\big)\times \bh(\bk) = 0$, defining $\bh_{\star\star}(\bk) = \bh(\bk)\hodge\big(\bh(\bk)\hodge \bh(\bk)\big)$, and the compact vector notation $\big(\text{tr}\big[\hat O\,\hat \blambda\big]\big)_a \equiv \text{tr}\big[\hat O\,\hat \lambda_a\big]$ \cite{Graf2021}. We find
\begin{align}
    \bb_n(\bk) = \frac{\Big[\big(E_n(\bk)-d_0(\bk)\big)^2-|\bh(\bk)|^2\Big]\bh(\bk)+\big(E_n(\bk)-d_0(\bk)\big)\,\bh_\star(\bk)+\bh_{\star\star}(\bk)}{\big(E_n(\bk)-d_0(\bk)\big)^3-\big(E_n(\bk)-d_0(\bk)\big)\,|\bh(\bk)|^2-\frac{1}{3}\bh(\bk)\cdot \bh_\star(\bk)}
    \label{eqn:bnVectorFormula}
\end{align}
enabling the calculation of the band projectors $\hat P_n(\bk)$ in terms of the corresponding band dispersion $E_n(\bk)$ and the Bloch Hamiltonian expanded in the {\it su}(4) generators via $d_0(\bk) = \frac{1}{4}\text{tr}\,\hat H(\bk)$ and $\bh(\bk) = \frac{1}{4}\text{tr}\big[\hat H(\bk)\,\hat \blambda\big]$.

\subsection{Incorporation of time-reversal and inversion symmetry}

\subsubsection{Time-reversal symmetry}

The anti-unitary time-reversal operator takes the form
\begin{align}
    \hat \cT = i\hat\sigma_2\hat\cK \equiv \hat 1\otimes i\hat\sigma_2\,\hat \cK \, ,
\end{align}
leaving the site invariant while reversing the spin components. The explicit action directly follows from Eqs.~\eqref{eqn:actionTx} to \eqref{eqn:actionTz}, leading to
\begin{align}
    &\{\hat 1, \hat \tau_1, \hat \tau_3, \hat \tau_2\hat \sigma_1, \hat \tau_2\hat \sigma_2, \hat \tau_2\hat \sigma_3\} \hspace{1cm} \hat \cT\text{-even} \, ,
    \label{eqn:TevenRepresentation}
\end{align}
while the remaining 10 combinations are $\cT$-odd. Note that $\hat \cK \hat\tau_2 \hat \cK = -\hat\tau_2$, while $\hat\tau_1$ and $\hat\tau_3$ are real.

\subsubsection{Representations of inversion operator}

We distinguish the inversion operators 
\begin{align}
    &\hat \cP_0 = \hat 1 \equiv \hat 1\otimes \hat 1 \,,\\
    &\hat \cP_x = \hat \tau_1 \equiv \hat\tau_1\otimes \hat 1 \,,\\
    &\hat \cP_z = \hat \tau_3 \equiv \hat\tau_3\otimes \hat 1 \,.
\end{align}
representing onsite inversion centers with the same $(\hat \cP_0)$ and different $(\hat\cP_z)$ parity and site exchange $(\hat \cP_x)$. In all cases, the spin part is left invariant. All are unitary satisfying $\hat \cP_a^2 = \hat 1_4$, such that $\hat \cP^{-1}_a = \hat \cP_a$. Concretely, the inversion operator $\hat \cP_0$ leaves all elements invariant, 
\begin{align}
    &\hat \cP_0\,\big(\hat \tau_a\hat \sigma_b\big) \,\hat \cP_0^{-1} = \hat \tau_a\hat \sigma_a \,
\end{align}
for $a=b=0,1,2,3$, while the inversion operator $\hat \cP_x$ acts as
\begin{align}
    &\hat \cP_x\,\hat \sigma_a\,\hat \cP_x^{-1} = \big(\hat \tau_1\hat \tau_1\big)\hat \sigma_a = \hat \sigma_a \, ,\\
    &\hat \cP_x\,\big(\hat \tau_1\hat \sigma_a\big)\,\hat \cP_x^{-1} = \big(\hat \tau_1\hat \tau_1\hat \tau_1\big)\hat \sigma_a = \hat \tau_1\hat \sigma_a \, ,\\
    &\hat \cP_x\,\big(\hat \tau_2\hat \sigma_a\big)\,\hat \cP_x^{-1} = \big(\hat \tau_1\hat \tau_2\hat \tau_1\big) \hat \sigma_a = -\hat \tau_2\hat \sigma_a \, ,\\
    &\hat \cP_x\,\big(\hat \tau_3\hat \sigma_a\big)\,\hat \cP_x^{-1} = \big(\hat \tau_1\hat \tau_3\hat \tau_1\big)\hat \sigma_a = -\hat \tau_3\hat \sigma_a \, ,
\end{align}
and the inversion operator $\hat \cP_z$ acts as
\begin{align}
    &\hat \cP_z\,\hat \sigma_a\,\hat \cP_z^{-1} = \big(\hat \tau_3\hat \tau_3\big)\hat \sigma_a = \hat \sigma_a \, ,\\
    &\hat \cP_z\,\big(\hat \tau_1\hat \sigma_a\big)\,\hat \cP_z^{-1} = \big(\hat \tau_3\hat \tau_1\hat \tau_3\big)\hat \sigma_a = -\hat \tau_1\hat \sigma_a \, ,\\
    &\hat \cP_z\,\big(\hat \tau_2\hat \sigma_a\big)\,\hat \cP_z^{-1} = \big(\hat \tau_3\hat \tau_2\hat \tau_3\big) \hat \sigma_a = -\hat \tau_2\hat \sigma_a \, ,\\
    &\hat \cP_z\,\big(\hat \tau_3\hat \sigma_a\big)\,\hat \cP_z^{-1} = \big(\hat \tau_3\hat \tau_3\hat \tau_3\big)\hat \sigma_a = \hat \tau_3\hat \sigma_a \, ,
\end{align}
for $a=0,1,2,3$.  

\subsubsection{Inversion-time-reversal symmetry}

The combined inversion-time-reversal symmetry reads
\begin{align}
    \hat \cP_0\hat \cT = i\hat \sigma_2\hat \cK\equiv \hat 1 \otimes i\hat \sigma_2\hat \cK \, , \hspace{1cm} \hat \cP_x\hat \cT = i\hat \tau_1\hat\sigma_2\hat \cK \equiv \hat \tau_1 \otimes i\hat \sigma_2\hat \cK \, , \hspace{1cm} \hat \cP_z\hat \cT = i\hat \tau_3\hat\sigma_2\hat\cK \equiv  \hat \tau_3 \otimes i\hat \sigma_2\hat \cK \, ,
\end{align}
leading to the separation into 
\begin{align}
    &\{\hat 1, \hat \tau_1, \hat \tau_3, \hat \tau_2\hat \sigma_1, \hat \tau_2\hat \sigma_2, \hat \tau_2\hat \sigma_3\} \hspace{1cm} \hat \cP_0\hat \cT\text{-even} \, ,
    \label{eqn:PTevenRepresentation1} \\
    &\{\hat 1, \hat \tau_1, \hat \tau_2, \hat \tau_3\hat \sigma_1, \hat \tau_3\hat \sigma_2, \hat \tau_3\hat \sigma_3\} \hspace{1cm} \hat \cP_x\hat \cT\text{-even} \, ,
    \label{eqn:PTevenRepresentation2} \\
    &\{\hat 1, \hat \tau_2, \hat \tau_3, \hat \tau_1\hat \sigma_1, \hat \tau_1\hat \sigma_2, \hat \tau_1\hat \sigma_3\} \hspace{1cm} \hat \cP_z\hat \cT\text{-even} \, ,
    \label{eqn:PTevenRepresentation3}
\end{align}
and 10 further $\hat\cP\hat\cT$-odd combinations.

\subsection{PT-even and T-even Dirac matrix representation}

We use the possibility to identify five $\hat\tau_a\otimes \hat \sigma_b$ denotes by $\hat \Gamma_i$~\cite{Kane2005, Fu2007} that satisfy
\begin{align}
    \big\{\hat \Gamma_i,\hat \Gamma_j\big\} = 2\,\delta_{ij}\,\hat 1_4
    \label{eqn:Clifford}
\end{align}
from which we construct the remaining ten matrices via their commutators
\begin{align}
    \hat \Gamma_{ij} = \frac{1}{2i}\big[\hat \Gamma_i,\hat \Gamma_j\big] \, .
    \label{eqn:CompletingClifford}
\end{align}
We use the convention to sort the basis elements as $\hat \blambda = (\hat \Gamma_1,...\,,\hat \Gamma_5, \hat\Gamma_{12}, ...\,, \hat \Gamma_{15}, \hat \Gamma_{23}, \hat \Gamma_{24},\hat \Gamma_{25}, \hat \Gamma_{34}, \hat\Gamma_{35}, \hat \Gamma_{45})$. We note that for all $i,j=1,...\,,5$, the symmetric structure constant vanishes, 
\begin{align}
    d_{ijc} = \frac{1}{8}\text{tr}\big(\{\hat \Gamma_i,\hat\Gamma_j\}\hat \lambda_c\big) = \frac{\delta_{ij}}{4}\,\text{tr}\,\hat \lambda_c = 0 \, .
\end{align}
For $\hat H(\bk) = d_0(\bk)\hat 1_4+\sum_{i=1}^5\,h_i(\bk)\,\Gamma_i$, we obtain $\bh_\star(\bk) = \bh_{\star\star}(\bk) = 0$, such that the remaining term in Eq.~\eqref{eqn:bnVectorFormula} simplifies to
\begin{align}
    \frac{\Big[\big(E_n(\bk)-d_0(\bk)\big)^2-|\bh(\bk)|^2\Big]\bh(\bk)}{\big(E_n(\bk)-d_0(\bk)\big)^3-\big(E_n(\bk)-d_0(\bk)\big)\,|\bh(\bk)|^2}  = \frac{\bh(\bk)}{E_n(\bk)-d_0(\bk)} \, ,
    \label{eqn:rank2Projector}
\end{align}
consistent with the rank-2 band projectors $\hat P_\pm(\bk) = \frac{1}{2}\big(\hat 1_4\pm \bh(\bk)/|\bh(\bk)|\cdot\hat \blambda\big)$ for double degenerate bands $E_n(\bk) = d_0(\bk)\pm |\bh(\bk)|$. We consider only cases where at least one $\hat \Gamma_{ij}$ breaks the band degeneracy and the full expression in Eq.~\eqref{eqn:bnVectorFormula} is required. We note that five traceless matrices in Eqs.~\eqref{eqn:PTevenRepresentation1}-\eqref{eqn:PTevenRepresentation3}, for instance the $\hat \cP_x\hat \cT$-even 
\begin{align}
    \hat \Gamma_1 = \hat \tau_1\,, \hspace{1cm} \hat \Gamma_2 = \hat \tau_2\,, \hspace{1cm} \hat \Gamma_3 = \hat \tau_3\hat \sigma_1\,, \hspace{1cm} \hat \Gamma_4 = \hat \tau_3\hat \sigma_2\,, \hspace{1cm} \hat \Gamma_5 = \hat \tau_3\hat \sigma_3\,, \hspace{1cm} \, ,
\end{align}
satisfy Eq.~\eqref{eqn:Clifford}. As a direct consequence of the definition in Eq.~\eqref{eqn:CompletingClifford}, $\hat \Gamma_{ij}$ are $\hat \cP_a\hat\cT$-odd. 

\subsection{Spin operator}

We note that 
\begin{align}
    \frac{1}{2i}\big[\hat \tau_a\hat\sigma_2,\hat \tau_a\hat\sigma_3\big] = \frac{1}{2i}\big[\hat \sigma_2,\hat\sigma_3\big] = \hat \sigma_1
\end{align}
for $a=1,2,3,$ and similarly for $\hat \sigma_1$ and $\hat\sigma_2$, such that we relate 
\begin{align}
    &\hat S_x = \hat \Gamma_{45}  \, ,\hspace{1cm} \hat S_y = -\hat \Gamma_{35} \, ,\hspace{1cm} \hat S_z = \hat \Gamma_{34} \, .
\end{align}
The relation holds both for all $\hat \cP_a\hat\cT$-even $\hat \Gamma_i$ in Eqs.~\eqref{eqn:PTevenRepresentation1}-\eqref{eqn:PTevenRepresentation3}. Let us denote the expansion of the spin operator for an arbitrary basis choice as
\begin{align}
    \hat S_a = \bar\be_a\cdot\hat \blambda \, .
    \label{eqn:spinOperatorBasis}
\end{align}
As seen by Eq.~\eqref{eqn:projectorBvector}, only a single component of $\bb_n(\bk)$ contributes to the spin polarization $\langle \hat S_a\rangle_n(\bk)$ within the basis choice $\hat \tau_i\hat\sigma_j$ in contrast to (generalized) Gellmann matrices~\cite{GellMann1962, Bertlmann2008, Graf2021}.

\section{Spin polarization of general four-band model}

Inserting Eqs.~\eqref{eqn:spinOperatorBasis} and \eqref{eqn:projectorBvector} for the spin and band projector, we obtain the spin polarization
\begin{align}
    \langle \hat S_a\rangle_{n}(\bk) = \text{tr}\big[\hat S_a\,\hat P_n(\bk)\big] = \frac{1}{4}\text{tr}\big[\big(\bar\be_a\cdot\hat \blambda\big)\big(\hat 1_4+\bb_n(\bk)\cdot\hat \blambda\big)\big] = \bar\be_a\cdot\bb_n(\bk) \, .
\end{align}
Using the expression for $\bb_n(\bk)$ as given in Eq.~\eqref{eqn:bnVectorFormula}, we obtain the spin polarization decomposed into three contributions
\begin{align}
    \langle \hat S_a\rangle_{n}(\bk) = \langle \hat S_a\rangle_{n,0}(\bk) + \langle \hat S_a\rangle_{n,\star}(\bk) + \langle \hat S_a\rangle_{n,\star\star}(\bk)
\end{align}
separated by the number of star products involved, 
\begin{align}
    &\langle \hat S_a\rangle_{n,0}(\bk) = c_{n,0}(\bk)\,\bar\be_a\cdot\bh(\bk) \, , \label{eqn:SpinGeneral1SM}\\ &\langle \hat S_a\rangle_{n,\star}(\bk) =c_{n,\star}(\bk)\,\bar\be_a\cdot\bh_\star(\bk)\, , \label{eqn:SpinGeneral2SM}\\&\langle \hat S_a\rangle_{n,\star\star}(\bk) =c_{n,\star\star}(\bk)\,\bar\be_a\cdot\bh_{\star\star}(\bk)\, . \label{eqn:SpinGeneral3SM}
\end{align}
The prefactors read
\begin{align}
    &c_{n,0}(\bk) = \frac{\big(E_n(\bk)-d_0(\bk)\big)^2-|\bh(\bk)|^2}{\big(E_n(\bk)-d_0(\bk)\big)^3-\big(E_n(\bk)-d_0(\bk)\big)\,|\bh(\bk)|^2-\frac{1}{3}\bh(\bk)\cdot \bh_\star(\bk)} \,, \label{eqn:prefactor1SM}\\[2mm]
    &c_{n,\star}(\bk) = \frac{E_n(\bk)-d_0(\bk)}{\big(E_n(\bk)-d_0(\bk)\big)^3-\big(E_n(\bk)-d_0(\bk)\big)\,|\bh(\bk)|^2-\frac{1}{3}\bh(\bk)\cdot \bh_\star(\bk)} \, , \label{eqn:prefactor2SM}\\[2mm]
    &c_{n,\star\star}(\bk) = \frac{1}{\big(E_n(\bk)-d_0(\bk)\big)^3-\big(E_n(\bk)-d_0(\bk)\big)\,|\bh(\bk)|^2-\frac{1}{3}\bh(\bk)\cdot \bh_\star(\bk)} \, . \label{eqn:prefactor3SM}
\end{align}

\subsection{Casimir invariants}

We aim to calculate the spin polarization $\langle \hat S_a\rangle_{n}(\bk)$ for the general four-band Hamiltonian in Eq.~\eqref{eqn:BlochHamiltonian}. For convenience, we choose the basis
\begin{align}
    \hat \blambda = \big(\hat\tau_1,\,\hat\tau_2,\,\hat\tau_3,\,\hat\bsigma,\,\hat\tau_1\hat\bsigma,\,\hat \tau_2\hat \bsigma,\,\hat \tau_3\hat\bsigma\big)
    \label{eqn:basisChoice}
\end{align}
for the traceless part, which yields 
\begin{align}
    \bh(\bk) = \big(d_1(\bk),\,d_2(\bk),\,d_3(\bk),\,\bd_0(\bk),\,\bd_1(\bk),\,\bd_2(\bk),\,\bd_3(\bk)\big) \, .
    \label{eqn:hamiltonianBasisExpansion}
\end{align}
In the following, we introduce the scalar and three-vector quantities $d_a(\bk)$ and $\bd_a(\bk)$ extended to the 15-dimensional space, which we denote by an overbar, e.g., $\bar d_1(\bk) \equiv \big(d_1(\bk), 0, ...,0\big)$. We obtain
\begin{align}
    &\frac{1}{4}C_2(\bk) = |\bh(\bk)|^2 = \sum_{a=1}^3d_a(\bk)^2+|\bd_0(\bk)|^2+\sum_{a=1}^3|\bd_a(\bk)|^2\, ,
\end{align}
relating the scalar product~\eqref{eqn:scalarSU} to the conventional 3-dimensional scalar product $|\bd_a(\bk)|^2=\bd_a(\bk)\cdot\bd_a(\bk)=d^1_a(\bk)^2+d^2_a(\bk)^2+d^3_a(\bk)^2$ for $\bd_a(\bk) = \big(d^1_a(\bk),d^2_a(\bk),d^3_a(\bk)\big)$. We find that the star product and the scalar product involving $\bar d_a(\bk)$ are related via
\begin{align}
    \bar d_a(\bk)\cdot \big[\bar \bd_0(\bk)\hodge \bar \bd_a(\bk)\big] = d_a(\bk)\,\bd_0(\bk)\cdot\bd_a(\bk) \, .
\end{align}
The star product involving three $\bd_a(\bk)$ relates to the volume element 
\begin{align}
    \bar \bd_1(\bk)\cdot\big[\bar \bd_2(\bk)\hodge \bar \bd_3(\bk)\big]= - \bd_1(\bk)\cdot\big[\bd_2(\bk)\times\bd_3(\bk)\big] \, .
\end{align}
These four combinations and their five symmetry-related counterparts arising from the cyclic property $\bn\cdot(\bM\hodge \bo) = \bo\cdot(\bn\hodge \bM)$ and the symmetry of the star product $\bn\hodge \bM = \bM\hodge \bn$ form the only nonzero contributions to $\bh(\bk)\cdot\bh_\star(\bk)$, leading to 
\begin{align}
    \frac{1}{4}C_3(\bk) = \bh(\bk)\cdot\bh_\star(\bk) = 6\,\sum_{a=1}^3d_a(\bk)\,\bd_0(\bk)\cdot\bd_a(\bk)-6\,\bd_1(\bk)\cdot\big[\bd_2(\bk)\times \bd_3(\bk)\big]\, .
\end{align}
We use the explicit result for $|\bh(\bk)|^2$ and $\bh(\bk)\cdot\bh_\star(\bk)$ to determine the prefactors~\eqref{eqn:prefactor1SM}-\eqref{eqn:prefactor3SM}. Further simplifications can be performed only for certain Hamiltonians, where a reasonable analytic expression for the band dispersions $E_n(\bk)$ is available.

\subsection{Vector structure of the spin polarization}

We reduce vector expressions in the 15-dimensional vector space to expressions in the 3-dimensional spatial space as follows. Using the basis choice~\eqref{eqn:basisChoice} and Bloch Hamiltonian vector~\eqref{eqn:hamiltonianBasisExpansion}, we read off
\begin{align}
    &\bar\be_a\cdot\bh(\bk) = \bd_0(\bk)\cdot\be_a \, ,
    \label{eqn:SpinPart1SM}
\end{align}
for Eq.~\eqref{eqn:SpinGeneral1SM}, indicating that $\langle \hat \bS\rangle_{n,0}(\bk)$ is parallel to $\bd_0(\bk)$ that captures the mean contributions of both sites A and B. We note that the only nonzero contributions to $\bar\be_a\cdot \bh_\star(\bk)$ in Eq.~\eqref{eqn:SpinGeneral1SM} arise from
\begin{align}
    &\bar\be_a\cdot\big[\bar d_b(\bk)\hodge\bar \bd_b(\bk)\big] = d_b(\bk)\,\bd_b(\bk)\cdot\be_a \, ,
\end{align}
for $a,b=1,2,3$ within the decomposition \eqref{eqn:hamiltonianBasisExpansion}, where $\bar d_b(\bk)$ and $\bar \bd_b(\bk)$ denotes the extensions to the 15-dimensional space. Including the symmetry of the star product, we obtain 
\begin{align}
    &\bar\be_a\cdot\bh_\star(\bk) = 2\sum_{b=1}^3d_b(\bk)\,\bd_b(\bk)\cdot\be_a \, ,
    \label{eqn:SpinPart2SM}
\end{align}
finding that $\langle \hat \bS\rangle_{n,\star}(\bk)$ is parallel to a superposition of spin-flip components $\bd_b(\bk)$ weighted by the respective spin-independent part $d_b(\bk)$. Similar contributions are also present in $\bar\be_a\cdot\bh_{\star\star}(\bk)$ in Eq.~\eqref{eqn:SpinGeneral3SM}, arising from 
\begin{align}
    &\bar\be_a\cdot\Big[\bar d_b(\bk)\hodge\big[\bar d_b(\bk)\hodge\bar\bd_0(\bk)\big]\Big] = d_b(\bk)^2 \,\bd_0(\bk)\cdot\be_a \, ,\\
    &\bar\be_a\cdot\Big[\bar \bd_b(\bk)\hodge\big[\bar\bd_b(\bk)\hodge\bar \bd_0(\bk)\big]\Big] = \bd_0(\bk)\cdot\bd_b(\bk)\,\bd_b(\bk)\cdot\be_a\, .
\end{align}
In addition, we find 
\begin{align}
    &\bar\be_a\cdot\Big[\bar d_1(\bk)\hodge\big[\bar\bd_2(\bk)\hodge\bar \bd_3(\bk)\big]\Big] = -d_1(\bk)\,\big(\bd_2(\bk)\times\bd_3(\bk)\big)\cdot\be_a\, , \\
    &\bar\be_a\cdot\Big[\bar d_2(\bk)\hodge\big[\bar\bd_3(\bk)\hodge\bar \bd_1(\bk)\big]\Big] = -d_2(\bk)\,\big(\bd_3(\bk)\times\bd_1(\bk)\big)\cdot\be_a\, , \\
    &\bar\be_a\cdot\Big[\bar d_3(\bk)\hodge\big[\bar \bd_1(\bk)\hodge\bar \bd_2(\bk)\big]\Big] = -d_3(\bk)\,\big(\bd_1(\bk)\times \bd_2(\bk)\big)\cdot\be_a\, , 
\end{align}
involving the cross product between $\bd_b(\bk)$. All remaining combinations vanish. Taking the combinatorial factors into account, the full expression reads
\begin{align}
    \bar\be_a\cdot\bh_{\star\star}(\bk) &=  2\sum_{b=1}^3\Big[d_b(\bk)^2\bd_0(\bk)+\bd_b(\bk)\cdot\bd_0(\bk)\,\bd_b(\bk)\Big]\cdot \be_a \nonumber \\
    &-2\Big[d_1(\bk)\,\big(\bd_2(\bk)\times\bd_3(\bk)\big)+d_2(\bk)\,\big(\bd_3(\bk)\times\bd_1(\bk)\big)+d_3(\bk)\,\big(\bd_1(\bk)\times \bd_2(\bk)\big)\Big]\cdot\be_a
    \label{eqn:SpinPart3SM}
\end{align}
not only yields contributions parallel to $\bd_0$ and $\bd_b$ but also orthogonal to the spin-flip contributions of the intersite coupling $\bd_b$.

\subsection{Examples}

In the following, we focus on examples, where $C_3(\bk) = 4\bh(\bk)\cdot \bh_\star(\bk) = 0$. This greatly simplifies the prefactors~\eqref{eqn:prefactor1SM}-\eqref{eqn:prefactor3SM}, reducing to
\begin{align}
    &c_{n,0}(\bk) = \frac{\big(E_n(\bk)-d_0(\bk)\big)^2-|\bh(\bk)|^2}{\big(E_n(\bk)-d_0(\bk)\big)^3-\big(E_n(\bk)-d_0(\bk)\big)\,|\bh(\bk)|^2} = \frac{1}{E_n(\bk)-d_0(\bk)}\,, \\[2mm]
    &c_{n,\star}(\bk) = \frac{E_n(\bk)-d_0(\bk)}{\big(E_n(\bk)-d_0(\bk)\big)^3-\big(E_n(\bk)-d_0(\bk)\big)\,|\bh(\bk)|^2} = \frac{1}{\big(E_n(\bk)-d_0(\bk)\big)^2-|\bh(\bk)|^2} \, , \\[2mm]
    &c_{n,\star\star}(\bk) = \frac{1}{\big(E_n(\bk)-d_0(\bk)\big)^3-\big(E_n(\bk)-d_0(\bk)\big)\,|\bh(\bk)|^2} = \frac{1}{E_n(\bk)-d_0(\bk)}\frac{1}{\big(E_n(\bk)-d_0(\bk)\big)^2-|\bh(\bk)|^2} \, . 
\end{align}
We focus on different classes of $\cP\cT$-symmetric models involving $d_0(\bk)$, $d_1(\bk)$, and $\bd_3(\bk)$ with $\cP\cT$-breaking perturbations, where analytic expressions for the band dispersions $E_n(\bk)$ exist. We summarize the results in Tab.~\ref{tab:SummaryExamplesSM}. A reduced version is given in Tab.~\ref{tab:OverviewSpinPolarizationGeneral} in the main text.

\subsubsection{Canted antiferromagnet}

We break the $\cP\cT$-symmetry by adding a term proportional to $\hat\sigma_a$, leading to the Hamiltonian
\begin{align}
    \hat H(\bk) = d_0(\bk)+d_1(\bk)\,\hat\tau_1+\bd_0(\bk)\cdot\hat\bsigma+\bd_3(\bk)\cdot\hat\tau_3\hat\bsigma \, .
\end{align}
The four (non-degenerate) bands read
\begin{align}
    E_{(ns)}(\bk) = d_0(\bk) + n\,\sqrt{d_1(\bk)^2+|\bd_0(\bk)|^2+|\bd_3(\bk)|^2+2\,s\,\sqrt{d_1(\bk)^2\,|\bd_0(\bk)|^2+(\bd_0(\bk)\cdot\bd_3(\bk))^2}}
\end{align}
with $n,s=\pm 1$, where we indicate the bands by increasing energy, that is, $E_1(\bk) = E_{(-+)}(\bk)$, $E_2(\bk)=E_{(--)}(\bk)$, $E_3(\bk)=E_{(+-)}(\bk)$, and $E_4(\bk)=E_{(++)}(\bk)$. Using the general expression in Eqs.~\eqref{eqn:SpinPart1SM}-\eqref{eqn:SpinPart3SM}, we obtain 
\begin{align}
    &\bar\be_a\cdot\bh(\bk) = \bd_0(\bk)\cdot\be_a \, , \\[0.5mm]
    &\bar\be_a\cdot\bh_\star(\bk) = 0 \, ,\\
    &\bar\be_a\cdot\bh_{\star\star}(\bk) =  2\big[d_1(\bk)^2\bd_0(\bk)+\bd_0(\bk)\cdot\bd_3(\bk)\,\bd_3(\bk)\big]\cdot \be_a \, ,
\end{align}
with corresponding prefactors
\begin{align}
    &c_{(ns),0}(\bk) = \frac{1}{E_{(ns)}(\bk)-d_0(\bk)} \, , \\
    &c_{(ns),\star\star}(\bk) = \frac{s}{2}\frac{1}{E_{(ns)}(\bk)-d_0(\bk)}\frac{1}{\sqrt{d_1(\bk)^2\,|\bd_0(\bk)|^2+(\bd_0(\bk)\cdot\bd_3(\bk))^2}} \, .
\end{align}
We find the spin polarization vector
\begin{align}
    \langle \hat \bS\rangle_{(ns)}(\bk) = \frac{1}{E_{(ns)}(\bk)-d_0(\bk)}\bigg[\bd_0(\bk)+s\,\frac{d_1(\bk)^2\,\bd_0(\bk)+\bd_0(\bk)\cdot\bd_3(\bk)\,\bd_3(\bk)}{\sqrt{d_1(\bk)^2\,|\bd_0(\bk)|^2+(\bd_0(\bk)\cdot\bd_3(\bk))^2}}\bigg] \, .
\end{align}

\subsubsection{Altermagnet}

We break the $\cP\cT$-symmetry by adding a term proportional to $\hat \tau_3$, leading to the Hamiltonian 
\begin{align}
    \hat H(\bk) = d_0(\bk)+d_1(\bk)\,\hat \tau_1+d_3(\bk)\,\hat \tau_3+\bd_3(\bk)\cdot\hat \tau_3\hat\bsigma
\end{align}
with (non-degenerate) bands 
\begin{align}
    &E_{(ns)}(\bk) = d_0(\bk) + n\,\sqrt{d_1(\bk)^2+d_3(\bk)^2+|\bd_3(\bk)|^2+2\,s\,|d_3(\bk)\,\bd_3(\bk)|}\, .
\end{align}
Using the general expression in Eqs.~\eqref{eqn:SpinPart1SM}-\eqref{eqn:SpinPart3SM}, we obtain 
\begin{align}
    &\bar\be_a\cdot\bh(\bk) =0 \, ,\\
    &\bar\be_a\cdot\bh_\star(\bk) = 2\,d_3(\bk)\,\bd_3(\bk)\cdot\be_a \, , \\
    &\bar\be_a\cdot\bh_{\star\star}(\bk) = 0 \, ,
\end{align}
with corresponding prefactor
\begin{align}
    &c_{(ns),\star}(\bk) = \frac{s}{2}\frac{1}{|d_3(\bk)\,\bd_3(\bk)|} \, .
\end{align}
We obtain the spin polarization vector
\begin{align}
    \langle \hat \bS\rangle_{(n\pm)}(\bk) =  \pm\,\text{sign}\big(d_3(\bk)\big)\,\frac{\bd_3(\bk)}{|\bd_3(\bk)|} \, .
\end{align}

\subsubsection{$P$-wave magnet}

We break the $\cP\cT$-symmetry by adding terms proportional to $\hat \tau_2\hat\sigma_a$, leading to the Hamiltonian
\begin{align}
    \hat H(\bk) = d_0(\bk)+d_1(\bk)\,\hat \tau_1+\bd_2(\bk)\cdot\hat \tau_2\hat\bsigma+\bd_3(\bk)\cdot\hat\tau_3\hat\bsigma
    \label{eqn:PwaveBlochHamiltonianGeneral}
\end{align}
with band dispersions
\begin{align}
    &E_{(ns)}(\bk) = d_0(\bk) + n\,\sqrt{d_1(\bk)^2+|\bd_2(\bk)|^2+|\bd_3(\bk)|^2+2\,s\,|\bd_2(\bk)\times\bd_3(\bk)|}\, .
    \label{eqn:PwaveBandStructureGeneral}
\end{align}
Using the general expression in Eqs.~\eqref{eqn:SpinPart1SM}-\eqref{eqn:SpinPart3SM}, we obtain 
\begin{align}
    &\bar\be_a\cdot\bh(\bk) = 0 \, , \\
    &\bar\be_a\cdot\bh_\star(\bk) = 0 \, , \\
    &\bar\be_a\cdot\bh_{\star\star}(\bk) = -2\,d_1(\bk)\,\big(\bd_2(\bk)\times\bd_3(\bk)\big)\cdot\be_a \, ,
\end{align}
with corresponding prefactors
\begin{align}
    &c_{(ns),\star\star}(\bk) = \frac{s}{2}\frac{1}{E_{(ns)}(\bk)-d_0(\bk)}\frac{1}{|\bd_2(\bk)\times\bd_3(\bk)|} \, .
\end{align}
We obtain the spin polarization vector
\begin{align}
    \langle \hat \bS\rangle_{(n\pm)}(\bk) = \mp\,\frac{d_1(\bk)}{E_{(n\pm)}(\bk)-d_0(\bk)}\frac{\bd_2(\bk)\times\bd_3(\bk)}{|\bd_2(\bk)\times\bd_3(\bk)|} \, .
    \label{eqn:SpinPwaveGeneralSM}
\end{align}

\subsubsection{Altermagnet with spin-orbit coupling}

We break the $\cP\cT$-symmetry by adding terms proportional to $\hat \tau_3$ and $\hat \tau_2\hat \sigma_a$, leading to the Hamiltonian
\begin{align}
    \hat H(\bk) = d_0(\bk)+d_1(\bk)\,\hat\tau_1+d_3(\bk)\,\hat\tau_3+\bd_2(\bk)\cdot\hat \tau_2\hat\bsigma+\bd_3(\bk)\cdot\hat \tau_3\hat\bsigma
\end{align}
with band dispersions
\begin{align}
    &E_{(ns)}(\bk) = d_0(\bk) + n\, \sqrt{d_1(\bk)^2+d_3(\bk)^2+|\bd_2(\bk)|^2+|\bd_3(\bk)|^2+2\,s\sqrt{d_3(\bk)^2\,|\bd_3(\bk)|^2+|\bd_2(\bk)\times\bd_3(\bk)|^2}}\, .
\end{align}
Using the general expression in Eqs.~\eqref{eqn:SpinPart1SM}-\eqref{eqn:SpinPart3SM}, we obtain 
\begin{align}
    &\bar\be_a\cdot\bh(\bk) = 0 \, , \\
    &\bar\be_a\cdot\bh_\star(\bk) = 2\,d_3(\bk)\,\bd_3(\bk)\cdot\be_a \, , \\
    &\bar\be_a\cdot\bh_{\star\star}(\bk) =  -2\,d_1(\bk)\,\big(\bd_2(\bk)\times\bd_3(\bk)\big)\cdot\be_a \, ,
\end{align}
with corresponding prefactors
\begin{align}
    &c_{(ns),\star}(\bk) = \frac{s}{2}\frac{1}{\sqrt{d_3(\bk)^2\,|\bd_3(\bk)|^2+|\bd_2(\bk)\times\bd_3(\bk)|^2}} \, , \\
    &c_{(ns),\star\star}(\bk) = \frac{s}{2}\frac{1}{E_{(ns)}(\bk)-d_0(\bk)}\frac{1}{\sqrt{d_3(\bk)^2\,|\bd_3(\bk)|^2+|\bd_2(\bk)\times\bd_3(\bk)|^2}} \, . 
\end{align}
We obtain the spin polarization vector
\begin{align}
    \langle \hat \bS\rangle_{(ns)}(\bk) = \pm\frac{1}{\sqrt{d_3(\bk)^2\,|\bd_3(\bk)|^2+|\bd_2(\bk)\times\bd_3(\bk)|^2}}\bigg[d_3(\bk)\,\bd_3(\bk)-\frac{d_1(\bk)}{E_{(ns)}(\bk)-d_0(\bk)}\,\bd_2(\bk)\times\bd_3(\bk)\bigg] \, .
\end{align}

\begin{table}[h!]
    \centering
    \begin{tabular}{cccccc}
        \hline\hline\\[-3mm]
         & &  Canted AFM & \hspace{3mm} AM \hspace{3mm} & \hspace{3mm} pWm\hspace{3mm}   & \hspace{5mm} AM with SOC \hspace{5mm} \\\hline\hline\\[-3mm]
        $\hat \tau_1$ & \hspace{6mm}$d_1(\bk)$ \hspace{5mm} & $\checkmark$ & $\checkmark$ & $\checkmark$ &  $\checkmark$\\
        $\hat \tau_2$& $d_2(\bk)$ &  \xmark & \xmark & \xmark  & \xmark \\
        $\hat \tau_3$& $d_3(\bk)$ &\xmark & $\checkmark$ & \xmark  & $\checkmark$\\
        $\hat \bsigma$& $\bd_0(\bk)$ &  $\checkmark$ & \xmark & \xmark  & \xmark \\
        $\hat \tau_1\hat \bsigma$& $\bd_1(\bk)$ &  \xmark & \xmark & \xmark  & \xmark \\
        $\hat \tau_2\hat \bsigma$& $\bd_2(\bk)$ & \xmark & \xmark & $\checkmark$  & $\checkmark$\\
        $\hat \tau_3\hat \bsigma$& $\bd_3(\bk)$ &  $\checkmark$ & $\checkmark$ & $\checkmark$  & $\checkmark$\\\hline\hline\\[-3mm]
        $\langle \hat \bS\rangle_{n,0}(\bk)$ & $\bM\cdot\bh(\bk)$ & $\bd_0(\bk)$ & $0$   & $0$ & $0$\\[1mm]\hline\\[-2mm]
        $\langle \hat \bS\rangle_{n,\star}(\bk)$ & $\bM\cdot\bh_{\star}(\bk)$ & $0$ &  $d_3(\bk)\,\bd_3(\bk)$  & $0$ & $d_3(\bk)\,\bd_3(\bk)$\\[1mm]\hline\\[-3mm]
        $\langle \hat \bS\rangle_{n,\star\star}(\bk)$ & $\bM\cdot\bh_{\star\star}(\bk)$&  \makecell{$d_1(\bk)^2\,\bd_0(\bk)$\\[1mm]$+\,\bd_0(\bk)\cdot\bd_3(\bk)\,\bd_3(\bk)$} & $0$   & $d_1(\bk)\,\bd_2(\bk)\times\bd_3(\bk)$& $d_1(\bk)\,\bd_2(\bk)\times \bd_3(\bk)$\\\\[-3mm]\hline\hline
    \end{tabular}
    \caption{Summary of general four-band Hamiltonian, for which analytic expressions of the band dispersions are available, which yield a physical interpretation as canted antiferromagnets (AFM), altermagnets (AM), $p$-wave magnet (pWm), altermagnet with spin-orbit coupling (SOC). We summarize the nonzero ($\checkmark$) and zero (\xmark) matrix components and the structure of the three contributions~\eqref{eqn:SpinGeneral1SM}-\eqref{eqn:SpinGeneral3SM}.}
    \label{tab:SummaryExamplesSM}
\end{table}

\section{Tight-binding model for p-wave magnet}

\subsection{Derivation of Bloch Hamiltonian}

Let us consider the two-dimensional tight-binding model for a p-wave magnet as presented in Ref.~\cite{Hellenes2024}. We derive the Bloch Hamiltonian from the real-space tight-binding model for completeness. Let us consider two sites A and B at spatial positions $\br_A = 0$ and $\br_B = \frac{1}{2}\be_x$, respectively, where $\be_x$ and $\be_y$ serve as coordinate system and lattice vectors within the x-y plane. We introduce the fermionic creation operators $\hat c^\dagger_{\bR,\nu} = (\hat c^\dagger_{\bR,\nu,\uparrow},\hat c^\dagger_{\bR,\nu,\downarrow})$ within unit cell $\bR$ for both sites $\nu=A,B$, where we align the coordinate system for the spin $\uparrow$ and $\downarrow$ with the spatial coordinate system extended to three dimensions via $\be_z$. We focus on nearest-neighbor hopping with real spin-independent hopping amplitude $t$ and spin- and site-dependent hopping amplitudes $\bJ_\nu\cdot\hat \bsigma$ that model the effect of exchange fields felt during the hopping process over in-plane magnetic moments located between the sites A and B. We distinguish $\bJ_x = J \big(\cos \varphi_x\,\be_x+\sin \varphi_x\,\be_y\big)$ and $\bJ_y = J \big(\cos \varphi_y\,\be_x+\sin \varphi_y\,\be_y\big)$ of equal amplitude but distinct spatial direction for hoppings in $\be_x$ and $\be_y$ characterized by the azimuthal angles $\varphi_x$ and $\varphi_y$ combined with a time-reversal-translation symmetry $\cT t_{\be_x/2}$ in x direction. The tight-binding Hamiltonian reads
\begin{align}
    \hat H &= \sum_\bR\,\hat c^\dagger_{B,\bR}\big[t+\bJ_x\cdot\hat\bsigma\big]\hat c^{}_{A,\bR}+\hat c^\dagger_{A,\bR+\be_x}\big[t-\bJ_x\cdot\hat\bsigma\big]\hat c^{}_{B,\bR}+\text{h.c.}\nonumber\\&+\sum_\bR\,\hat c^\dagger_{A,\bR+\be_y}\big[t+\bJ_y\cdot\hat\bsigma\big]\hat c^{}_{A,\bR}+\hat c^\dagger_{B,\bR+\be_y}\big[t-\bJ_y\cdot\hat\bsigma\big]\hat c^{}_{B,\bR}+\text{h.c.} \, ,
\end{align}
where h.c. denotes Hermitian conjugation. Performing the Fourier transformation via
\begin{align}
    \hat c^\dagger_{\bR,\nu}=\frac{1}{\sqrt{N_c}}\sum_\bk\,e^{-i\bk\cdot(\bR+\br_\alpha)}\,\hat c^\dagger_{\bk,\nu}
\end{align}
where $N_c$ denotes the number of unit cells, we obtain $\hat H = \sum_\bk \hat c^\dagger_\bk \,\hat H(\bk)\,\hat c^{}_\bk$ within the basis $\hat c^\dagger_\bk = \big(\hat c^\dagger_{A\uparrow}(\bk),\hat c^\dagger_{A\downarrow}(\bk),\hat c^\dagger_{B\uparrow}(\bk),\hat c^\dagger_{B\downarrow}(\bk)\big)$ taking the form
\begin{align}
    \hat H(\bk) =  \hat 1_2\otimes\big(2t\cos k_y \hat 1_2\big)+\hat \tau_z\otimes \big(2\cos k_y\,\bJ_y\cdot \hat\bsigma\big)+\hat\tau_x\otimes\big(2t\cos\frac{k_x}{2}\hat 1_2\big)+\tau_y\otimes\big(-2\sin \frac{k_x}{2}\,\bJ_x\cdot\hat\bsigma\big) \, ,
\end{align}
which we give in Eq.~\eqref{eqn:PwaveBlochHamiltonian} in the main text. Taking the general form given in Eq.~\eqref{eqn:BlochHamiltonian} in the main text, we read off
\begin{align}
    d_0(\bk) = 2t\cos k_y \, , \hspace{1cm}
    d_1(\bk) = 2t \cos\frac{k_x}{2} \, , \hspace{1cm}
    \bd_2(\bk) = -2\sin \frac{k_x}{2}\,\bJ_x \, , \hspace{1cm}  
    \bd_3(\bk) = 2\cos k_y\,\bJ_y    \, .
    \label{eqn:dVecPwave}
\end{align}
The Hamiltonian is of the form given in Eq.~\eqref{eqn:PwaveBlochHamiltonianGeneral}. We give a summary in Tab.~\ref{tab:SummaryMinimalModelSM}. In the following, we study the dependence of the spin splitting and spin polarization on the amplitude $J/t$ and the relative direction $\varphi=\varphi_x-\varphi_y$ of the in-plane magnetic moments.

\subsection{Band gap}

The general band dispersion for four-band Hamiltonians involving $\hat \tau_1$, $\hat\tau_2\hat\bsigma$, and $\hat\tau_3\hat\bsigma$ is given in Eq.~\eqref{eqn:PwaveBandStructureGeneral} involving the cross product
\begin{align}
    \bd_2(\bk)\times \bd_3(\bk)\propto \hat \bJ_x \times \hat \bJ_y = \big(\cos \varphi_x\,\sin\varphi_y-\sin\varphi_x\cos\varphi_y\big)\,\be_z= -\sin(\varphi_x-\varphi_y)\,\be_z \, .
\end{align}
As expected, only the relative angle between the local magnetic moments $\varphi=\varphi_x-\varphi_y$ is relevant. We obtain the band dispersions
\begin{align}
    &E_{(ns)}(\bk) = 2t\cos k_y + 2|t|\,n\,\sqrt{\cos^2(k_x/2)+\Big(\frac{J}{t}\Big)^2\Big[\sin^2 (k_x/2)+\cos^2(k_y)+2\,s\,|\sin(k_x/2)\cos (k_y)\sin\varphi |\Big]}\, .
\end{align}
The spin splitting between the two lowest bands $E_1(\bk) = E_{(-+)}(\bk)$ and $E_2(\bk) = E_{(--)}(\bk)$ read to leading order in momentum is
\begin{align}
    \Delta_{12}(\bk) = E_2(\bk)-E_1(\bk) \approx \frac{2J^2}{\sqrt{t^2+J^2}}|k_x\,\sin\varphi| \, .
\end{align}
The band gap to the next band $E_3(\bk) = E_{(+-)}(\bk)$ is to leading order
\begin{align}
    \Delta_{13}(\bk) = E_3(\bk)-E_1(\bk) \approx 4\sqrt{t^2+J^2} \, .
\end{align}

\begin{table}
    \centering
    \begin{tabular}{c|cc|c|c|ccc|c|cc}
        \hline\hline
        & $\cP_x$ & $\cT$ &  $\cP_x\cT$-even/odd basis &  &$p$-wave magnet~\cite{Hellenes2024} \\ \hline
        $\hat 1$     &$+$& $+$ & - & $d_0(\bk)$   &   $2t\cos k_y$   \\
        $\hat \bsigma$ &$+$&$-$& $(\hat\Gamma_{45},-\hat\Gamma_{35},\hat\Gamma_{34})$  & $\bd_0(\bk)$ &  -      \\ \hline
        $\hat \tau_1$       &$+$&$+$& $\hat\Gamma_{1}$ & $d_1(\bk)$    &  $2t\cos(k_x/2)$     \\
        $\hat\tau_1\hat \bsigma$   &$+$&$-$   & $(\hat\Gamma_{23},\hat\Gamma_{24},\hat\Gamma_{25})$    & $\bd_1(\bk)$                                   &  -                                                          \\ \hline
        $\hat\tau_2$       &$-$&$-$                                       & $\hat\Gamma_{2}$                                     &  $d_2(\bk)$                                     &-                                                           \\
        $\hat\tau_2\hat \bsigma$   &$-$&$+$         & $(-\hat\Gamma_{13},-\hat\Gamma_{14},-\hat\Gamma_{15})$  & $\bd_2(\bk)$ &                                 $-2J\sin(k_x/2)\,\hat\bJ_x$                            \\
        \hline
        $\hat\tau_3$       &$-$&$+$                                    & $\hat\Gamma_{12}$     & $d_3(\bk)$& -                                                                     \\
        $\hat\tau_3\hat \bsigma$  &$-$&$-$   & $(\hat\Gamma_{3},\hat\Gamma_{4},\hat\Gamma_{5})$   &  $\bd_3(\bk)$    &  $2J\cos(k_y)\,\hat\bJ_y$                             \\ \hline\hline
    \end{tabular}
    \caption{Overview of the matrix decomposition of a general four-band Hamiltonian and their transformation under inversion $\hat\cP_x = \tau_1$ and time-reversal transformation $\cT$. We give the corresponding $\cP_x\cT$-even $\hat\Gamma_i$ and $\cP_x\cT$-odd $\hat\Gamma_{ij}$ with the corresponding coefficients in the general case and for the minimal model for a $p$-wave magnet. A reduced and differently arranged version is given in Tab.~\ref{tab:SummaryMinimalModelPwaveMainText} in the main text. }
    \label{tab:SummaryMinimalModelSM}
\end{table}

\subsection{Spin polarization}

Inserting \eqref{eqn:dVecPwave} into the spin polarization in Eq.~\eqref{eqn:SpinPwaveGeneralSM} yields
\begin{align}
    \frac{\bd_2(\bk)\times\bd_3(\bk)}{|\bd_2(\bk)\times\bd_3(\bk)|} = -\frac{\sin(k_x/2)\,\cos k_y}{|\sin(k_x/2)\,\cos k_y|}\frac{\hat \bJ_x\times\hat \bJ_y}{|\hat \bJ_x\times\hat \bJ_y|} = \text{sign}\big[\sin(k_x/2)\,\cos k_y\,\sin\varphi\big]\,\be_z \, ,
\end{align}
pointing the out-of-plane direction $\be_z$, where the momentum and the chirality of the in-plane magnetic moments give the sign. The amplitudes for the four bands labeled by $n,s=\pm$ are given by 
\begin{align}
    S_{(ns)}(\bk)&=-s\frac{d_1(\bk)}{E_{(ns)}(\bk)-d_0(\bk)} \\
    &=-n\,s\frac{\text{sign}[t]\cos(k_x/2)}{\sqrt{\cos^2(k_x/2)+\Big(\frac{J}{t}\Big)^2\Big[\sin^2 (k_x/2)+\cos^2(k_y)+2\,s\,|\sin(k_x/2)\cos (k_y)\sin\varphi |\Big]}} \, .
    \label{eqn:PolarizationAmplitudeSM}
\end{align}
The amplitude simplifies to 
\begin{align}
    S_{(ns)}(\bk)\approx-n\,s\,\text{sign}\big[t\,\cos(k_x/2)\big] \hspace{1cm} \text{for}\hspace{1cm} J\ll t \, ,
\end{align}
leading to fully spin polarized bands in leading order of $J/t$, i.e.,
\begin{align}
    &\langle \hat \bS\rangle_{1}(\bk) \approx \text{sign}\big[t\,\sin(k_x/2)\,\cos(k_x/2)\,\cos k_y\,\sin\varphi\big]\be_z \approx \text{sign}\big[t\,k_x\,\sin\varphi\big]\be_z \, , \\ 
    &\langle \hat \bS\rangle_{2}(\bk) \approx -\text{sign}\big[t\,\sin(k_x/2)\,\cos(k_x/2)\,\cos k_y\,\sin\varphi\big]\be_z  \approx -\text{sign}\big[t\,k_x\,\sin\varphi\big]\be_z \, , \\
    &\langle \hat \bS\rangle_{3}(\bk) \approx \text{sign}\big[t\,\sin(k_x/2)\,\cos(k_x/2)\,\cos k_y\,\sin\varphi\big]\be_z  \approx \text{sign}\big[t\,k_x\,\sin\varphi\big]\be_z \, , \\
    &\langle \hat \bS\rangle_{4}(\bk) \approx -\text{sign}\big[t\,\sin(k_x/2)\,\cos(k_x/2)\,\cos k_y\,\sin\varphi\big]\be_z  \approx -\text{sign}\big[t\,k_x\,\sin\varphi\big]\be_z \, .
\end{align}
where we expanded for small momenta in the second step. Expanding Eq.~\eqref{eqn:PolarizationAmplitudeSM} for small momenta, we obtain to first order
\begin{align}
    S_{(ns)}(\bk) \approx -n\frac{t}{\sqrt{t^2+J^2}}\bigg[s-\frac{1}{2}\frac{J^2}{t^2+J^2}|k_x\,\sin\varphi|\bigg] \, ,
\end{align}
leading to the spin polarization
\begin{align}
    &\langle \hat \bS\rangle_{1}(\bk) \approx  \text{sign}\big[k_x\,\sin\varphi\big]\frac{t}{\sqrt{t^2+J^2}}\bigg[1-\frac{1}{2}\frac{J^2}{t^2+J^2}|k_x\,\sin\varphi|\bigg]\be_z\, , \\ 
    &\langle \hat \bS\rangle_{2}(\bk) \approx  -\text{sign}\big[k_x\,\sin\varphi\big]\frac{t}{\sqrt{t^2+J^2}}\bigg[1+\frac{1}{2}\frac{J^2}{t^2+J^2}|k_x\,\sin\varphi|\bigg]\be_z\, , \\
    &\langle \hat \bS\rangle_{3}(\bk) \approx  \text{sign}\big[k_x\,\sin\varphi\big]\frac{t}{\sqrt{t^2+J^2}}\bigg[1+\frac{1}{2}\frac{J^2}{t^2+J^2}|k_x\,\sin\varphi|\bigg]\be_z\, , \\
    &\langle \hat \bS\rangle_{4}(\bk) \approx  -\text{sign}\big[k_x\,\sin\varphi\big]\frac{t}{\sqrt{t^2+J^2}}\bigg[1-\frac{1}{2}\frac{J^2}{t^2+J^2}|k_x\,\sin\varphi|\bigg]\be_z\, .
\end{align}

\section{ab initio method}

Ab initio calculations were performed using density functional theory (DFT). We used the \textsc{VASP} DFT code~\cite{kresse_ab_1993, kresse_ab_1994, kresse_efficiency_1996, kresse_efficient_1996}, which employs the projector-augmented-wave (PAW) method \cite{blochl_projector_1994, kresse_ultrasoft_1999}. 
Perdew-Burke-Ernzerhof (PBE) functional \cite{perdew_generalized_1996, perdew_generalized_1997} with the generalized-gradient approximation was used to model exchange-correlation effects. We used the recommended pseudopotentials for Ce, Ni, As, and O from the standard collection provided within \textsc{VASP}.

The plane-wave energy cutoff was set to 800 eV, and a full $6 \times 8 \times 4$ k-points $\Gamma$-centered grid (without any symmetry reductions) was used in self-consistent calculations. The energy convergence criterion was set to $10^{-7}$ eV. The magnetic configurations were stabilized using the constrained local moments approach available in \textsc{VASP} \cite{ma_constrained_2015, hobbs_fully_2000}.

\end{widetext}

\end{document}